\documentclass[traditabstract]{aa} 
\usepackage{graphicx}
\usepackage{txfonts}
\usepackage{longtable}
\usepackage[authoryear]{natbib}
\bibpunct{(}{)}{;}{a}{}{,}

\def\ltsima{$\; \buildrel < \over \sim \;$}
\def\simlt{\lower.5ex\hbox{\ltsima}}
\def\gtsima{$\; \buildrel > \over \sim \;$}
\def\simgt{\lower.5ex\hbox{\gtsima}}
\newcommand{\HI}{\ion{H}{i}}

\begin{document}
   \title{The extended structure of the dwarf irregular galaxy Sagittarius\thanks{Based on data obtained with the European Southern Observatory Very Large Telescope, Paranal, Chile, under the program 089.D-0052(A).},\thanks{Table of stellar photometry is only available at the Strasbourg astronomical Data Center (CDS) via anonymous ftp to cdsarc.u-strasbg.fr (130.79.128.5)
or via http://cdsweb.u-strasbg.fr/cgi-bin/qcat?J/A+A/}}

   \author{G. Beccari\inst{1}, M. Bellazzini\inst{2}, F. Fraternali\inst{3}, G. Battaglia\inst{2,8}, S. Perina\inst{4}, A. Sollima\inst{2}, T.A. Oosterloo\inst{5,6}, V. Testa\inst{7},  
           \and S. Galleti\inst{2}
           }
         
      \offprints{M. Bellazzini}

   \institute{European Southern Observatory, Alonso de Cordova 3107, Vitacura Santiago, Chile. 
              \email{gbeccari@eso.org} 
           \and
             INAF - Osservatorio Astronomico di Bologna,
              Via Ranzani 1, 40127 Bologna, Italy
           \and
              Dipartimento di Astronomia - Universit\`a degli Studi di Bologna,
            Via Ranzani 1, 40127 Bologna, Italy
            \and
           Department of Astronomy and Astrophysics, Pontificia Universidad Cat\'olica de Chile, Avenida Vicu\~na Mackenna 4860, Macul, Santiago, Chile 
            \and
            Netherlands Institute for Radio Astronomy, Postbus 2, 7990 AA Dwingeloo, the Netherlands
            \and
            Kapteyn Astronomical Institute, University of Groningen, Postbus 800, 9700 AV Groningen, 
            The Netherlands
            \and
             INAF - Osservatorio Astronomico di Roma, via Frascati 33, 00040 Monteporzio, Italy
            \and
            Instituto de Astrof'sica de Canarias, 38205 La Laguna, Tenerife, Spain
           }

     \authorrunning{G. Beccari et al.}
   \titlerunning{The extended structure of the dwarf irregular galaxy Sagittarius}

   \date{Accepted for publication by A\&A }

\abstract{We present a detailed study of the stellar and \HI\ structure of the dwarf irregular galaxy Sagittarius. We use new deep and wide field photometry to trace the surface brightness profile of the galaxy out to $\simeq 5.0\arcmin$ (corresponding to $\simeq 1600$~pc) and down to $\mu_V\simeq 30.0$~mag/arcsec$^2$, thus showing that the stellar body of the galaxy is much more extended than previously believed, and it is similarly (or more) extended than the overall \HI\ distribution. The whole major-axis profile is consistent with a pure exponential, with a scale radius of $\simeq 340$~pc. The surface density maps reveal that the distribution of old and intermediate-age stars is smooth and remarkably flattened out to its edges, while the associated \HI\ has a much rounder shape, is off-centred and presents multiple density maxima and a significant hole. No clear sign of systemic rotation is detectable in the complex \HI\ velocity field. 
No metallicity gradient is detected in the old and intermediate age population of the galaxy, and we confirm that this population has a much more extended distribution than young stars (age$\la 1$~Gyr).  }


\maketitle
%

\section{Introduction}
\label{intro}

The origin of the variety of morphologies of dwarf galaxies and the evolutionary path leading to the formation of the amorphous and gas-devoid dwarf Spheroidal (dSph) galaxies have important consequences for our understanding of galaxy formation and cosmology \citep{mateo,tht}. The well-known fact that dSphs are found in the surroundings of large galaxies while gas-rich dwarf Irregulars - dIrr - are farther away, on average,
\citep[i.e., the local morphology-density relation, see, e.g.,][]{grce} has led to the idea that strong interactions with the large galaxies that they are orbiting around have transformed primordial actively star-forming dIrrs into present-day quiescent dSph, removing the gas and stopping the star formation \citep{mateo}. 

Recent state-of-the-art modelling in cosmological context has revealed that tidal interactions and ram-pressure stripping can be quite efficient in transforming gas-rich disc dwarfs (plunging deeply into the halo of their main galaxy) into dSphs \citep[the tidal stirring model,][]{lucionat}, but also supernova (SN) winds and the cosmic ultraviolet (UV) background are also found to play significant roles in removing gas from dwarfs \citep{kaza,saw10,saw12}. Observations of isolated dwarfs in the Local Group (LG) are expected to provide  crucial insight into the impact of the various factors on shaping the current status of dSphs, since they may represent the test case of evolution without interactions and can give insight into the initial conditions at the epoch of formation. These considerations were the driving case for a large Hubble Space Telescope (HST) programme \citep{lcid} that performs extremely deep observations in small central fields of a small sample of isolated galaxies, searching for the effects of SN feedback and/or cosmic re-ionisation, since they may be recorded in the star formation history (SFH) of the systems \citep[see, e.g.,][]{monel_cet,monel_tuc}. 

We have recently started a research project aimed at exploring a fully complementary aspect, i.e., the large scale structure of these galaxies, both of their stellar body and of their gas (\HI). The lack of tidal limits imposed by nearby stellar systems implies that isolated galaxies may preserve extended stellar haloes at low SB, as it has been indeed observed in some  cases (e.g., \citealt{vanse}, in Leo A, or \citealt{sanna}, in IC10). This kind of feeble but extended structure (a) records crucial information on the formation and evolution of these galaxies and (b) provides the precious stellar test particles to study the dynamics of these systems at large distances from their centres, thus probing their dark matter (DM) haloes over a wide radial range. The structure and kinematics of the neutral hydrogen provide additional and complementary information. It must be stressed that dwarf galaxies evolved in isolation should have preserved their pristine DM haloes virtually untouched, hence bearing fundamental information on their original structure and mass distribution.

In previous papers of this series \citep[][hereafter Papers I and II, respectively]{pap1,pap2},
we demonstrated that the outer structures of these galaxies can be traced down to extremely low surface brightness (SB) levels ($\mu_V\simeq 30$ mag/arcsec$^2$) with star counts from very deep photometry obtained with state of the art wide-field cameras on 8~m class telescopes, with modest amounts of observing time, under excellent seeing conditions. In Papers I and II we studied targets visible from the northern hemisphere, that have been observed at the Large Binocular Telescope (LBT), using the Large Binocular Camera \citep[LBC, ][]{lbc}. 

In this paper we present the results of the same kind of analysis as performed with the VIMOS camera at the ESO Very Large Telescope (VLT) of our first southern target, the dwarf irregular galaxy Sagittarius (Sgr~dIrr 
\footnote{Other names are UKS~1927-177 and ESO 594-4. This galaxy has been sometime denoted as Sag-DIG. This name is highly unconventional since the official abbreviation for the constellation of Sagittarius is {\em Sgr}, not Sag, and the classical abbreviation for dwarf irregular is dIrr, not DIG. For this reason, we think that the name `Sag-DIG` should be dismissed.}). 
Sgr~dIrr  is a faint star-forming galaxy \citep[see][hereafter M05]{moma} which is among the most gas-rich dwarfs in the local volume \citep{grce}. It is more distant than 1~Mpc from both the MW and M31. It was independently discovered by \citet{cesa} and \citet{long} on photographic plates. It is located right at the edge of the LG \citep[][M12 hereafter]{mcc}, and according to \citet{teys12}, it is likely to be falling into the LG for the first time. The most recent and detailed study of the stellar content and star formation history (SFH) of Sgr~dIrr is that by M05, from data obtained with the Advanced  Camera for Surveys (ACS) onboard the Hubble Space Telescope (HST). Star formation has been taking place in this galaxy since $\sim 10$~Gyr ago (M05), with a possible significant enhancement in the star formation rate that occurred between $\sim 7$ and $\sim 3$~Gyr ago \citep{weisz}. 

The structure of the paper is the following. In Sect.~\ref{phot} we present the observational material and we
describe the data reductions and calibrations, we define the final photometric samples and present the color-magnitude diagrams (CMD);  in Sect.~\ref{grad} we search for metallicity gradients among old stars and briefly
consider the spatial distribution of the various stellar species. In Sect.~\ref{struc} we present the SB
profile and density maps. In Sect.~\ref{HI} we reconsider the structure and dynamics of the \HI\
distributions associated to Sgr~dIrr, and finally, in Sect.~\ref{disc} we summarise and discuss our results. 

\section{Observations and data reduction}
\label{phot}

Our photometric data were acquired during different nights between May 28 and July 18, 2012, using the recently refurbished VIMOS, a multi-object spectrograph and wide-field imager mounted on the Nasmyth focus B of the ESO Very Large Telescope (VLT) UT3 (Melipal). The camera is  composed of four identical 
4000~px $\times$ 2000~px CCDs, each covering a field of $7\arcmin\times8\arcmin$ (pixel scale $0.205\arcsec$/px), with a gap of $\sim 2\arcmin$ between each quadrant. 
Observations were obtained in four different pointings (P1, P2, P3, and P4, hereafter) following the scheme adopted by \citet{matte} to mosaic a continuous field (i.e., without gaps) of 
$\ga 20\arcmin\times 20\arcmin$. In such a scheme the centre of the galaxy is imaged in one chip in all the pointings, so that the deepest and more accurate photometry can be obtained in the central part of the galaxy. 
The average seeing in the four pointings was 1.3$\arcsec$, 1.5$\arcsec$, 1.1$\arcsec$, and 1.1$\arcsec$  FWHM, from P1 to P4, respectively. For each pointing, six to eight $t=255$~s exposures per filter were acquired, in V and I. 

   \begin{figure}
   \centering
   \includegraphics[width=\columnwidth]{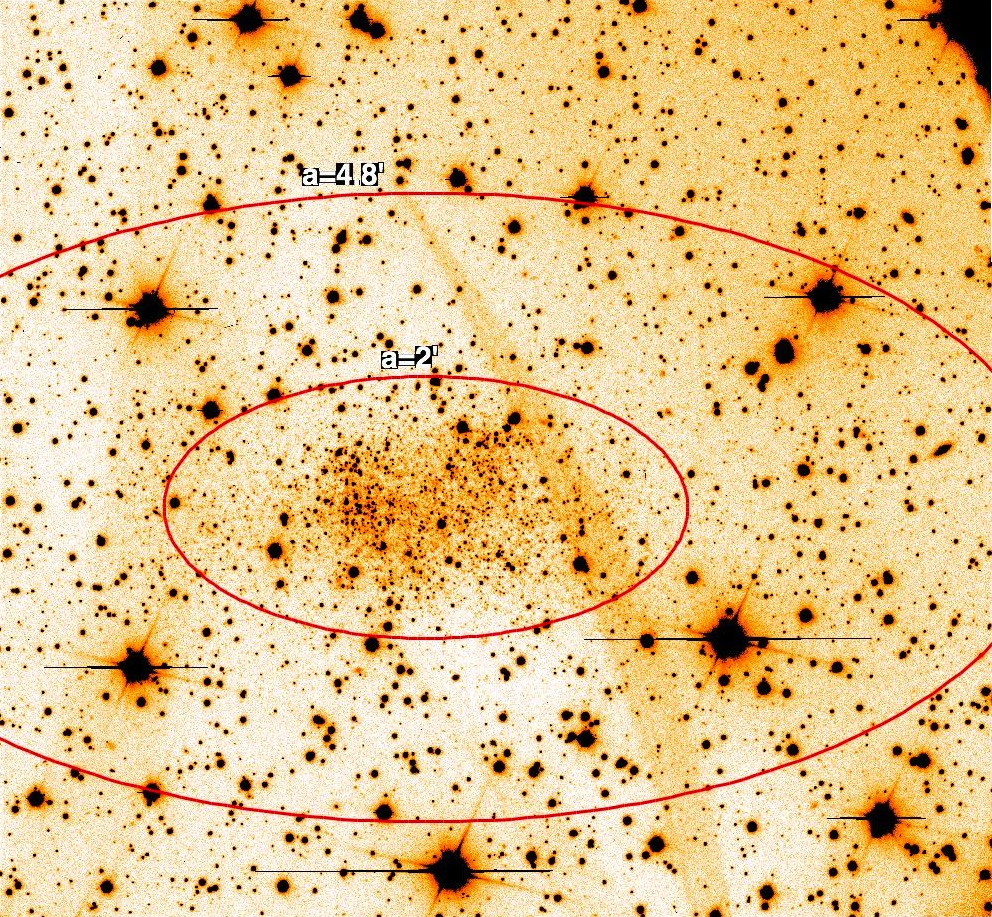}
     \caption{Deep V band images of a VIMOS chip nearly centred on Sgr~dIrr. North is up, east to the left. The two superimposed ellipses are centred on the centre of the galaxy and have the same ellipticity and PA assumed for the galaxy (see Table~\ref{Tab_par}), and have semi-major axis (a) of $2\arcmin$ and $4.8\arcmin$.}
        \label{imaC}
    \end{figure}


Relative photometry was performed independently on each image using the PSF-fitting code DAOPHOTII/ALLSTAR \citep{daophot,allframe}. Sources with a peak higher than 3$\sigma$ above the background were identified in a stacked image obtained by registering and co-adding all the images considered for the analysis. Then, each of these stars was re-identified and fitted on each image, when possible.  To clean the catalogues  from spurious sources, we also adopted cuts in the image quality parameters $CHI$ and $SHARP$, provided by DAOPHOTII. After accurate inspection of the distribution of measured sources in the planes mag vs. $CHI$ and mag vs. $SHARP$, we decided to retain only those sources having $CHI \le 2.0$ and SHARP within magnitude-dependent contours traced to include the bulk of stellar objects, as done in Paper~II. 

We combined the raw photometry in two ways to obtain two different catalogues/samples:

\begin{enumerate}

\item The MOSAIC sample, which combines all the photometry from the four pointings to cover a field of $\simeq 26\arcmin\times22\arcmin$ as uniformly as possible. 

\item The IN6 sample, which combines the repeated observations of the centre of Sgr~dIrr into a field of 
$\simeq 10\arcmin\times 8\arcmin$ with the best relative photometry and only contains sources  having at least {\em six} independent magnitude estimates per filter.

\end{enumerate} 

The sky coverage of the two samples is illustrated in Fig.~\ref{mappet}. In the following subsections we describe in detail how these two samples have been constructed.

\subsection{The MOSAIC sample}

In the final catalogue of each individual chip, we retained only the sources with at least three independent magnitude estimates per filter. The average and the standard error of the mean of the independent measures obtained from the different images were adopted as the final values of the instrumental magnitude and of the uncertainty on the relative photometry. We independently transformed the x,y coordinates (in pixels) of each of these 16 subsamples (4~chips $\times$ 4~pointings = 16~chips)
into J2000 Equatorial coordinates with second degree polynomials. The transformations were obtained by model fitting to 300-800 stars in common between each chip and the GSC2.2\footnote{The Guide Star Catalogue-II is a joint project of the Space Telescope Science Institute and the Osservatorio Astronomico di Torino. Space Telescope Science Institute is operated by the Association of Universities for Research in Astronomy, for the National Aeronautics and Space Administration under contract NAS5-26555. The participation of the Osservatorio Astronomico di Torino is supported by the Italian Council for Research in Astronomy. Additional support is provided by European Southern Observatory, Space Telescope European Coordinating Facility, the International GEMINI project and the European Space Agency Astrophysics Division.} catalogue of astrometric standards, with a typical r.m.s of $\le 0.25\arcsec$ in each coordinate\footnote{All the astrometric solutions presented here have been obtained with CataXcorr, a code  developed by P. Montegriffo at INAF - Osservatorio Astronomico di Bologna, and successfully used by our group during the past 10 years. See {\tt \tiny http://davide2.bo.astro.it/~paolo/Main/CataPack.html}}.
Then all the individual catalogues were moved into a common relative photometry system using the stars in common in the overlapping areas between chips of different pointings. Finally all the catalogues of the individual chips were merged. Magnitudes and positions of stars in common between different chips were taken from the best dataset, adopting the following hierarchy: P3, P4, P1, P2. In this way each star in the MOSAIC catalogue has its magnitudes (and associated errors) obtained from three independent estimates per filter, independently of its position in the field, thus ensuring the highest level of homogeneity attainable over the widest field that can be covered with our data. Inspection of Fig.~\ref{mappet} reveals that the various degrees of overlap between different chips, the variations in mean seeing between different pointings, and the presence of heavily saturated bright foreground stars in the field introduces some degree of inhomogeneity in the MOSAIC sample. Still this is the most uniform sample on the widest field that we can obtain from the available observational material. We refined the global astrometry of the MOSAIC catalogue as a whole with a fourth degree polynomial fitted to the 7042 stars in common with GSC2.2. The final r.m.s is $\simeq 0.23\arcsec$ in each coordinate. We use the MOSAIC sample in Sect.~\ref{struc} to trace the full extension of the stellar body of Sgr~dIrr.

\subsection{The IN6 sample}

\begin{figure}
   \centering
   \includegraphics[width=\columnwidth]{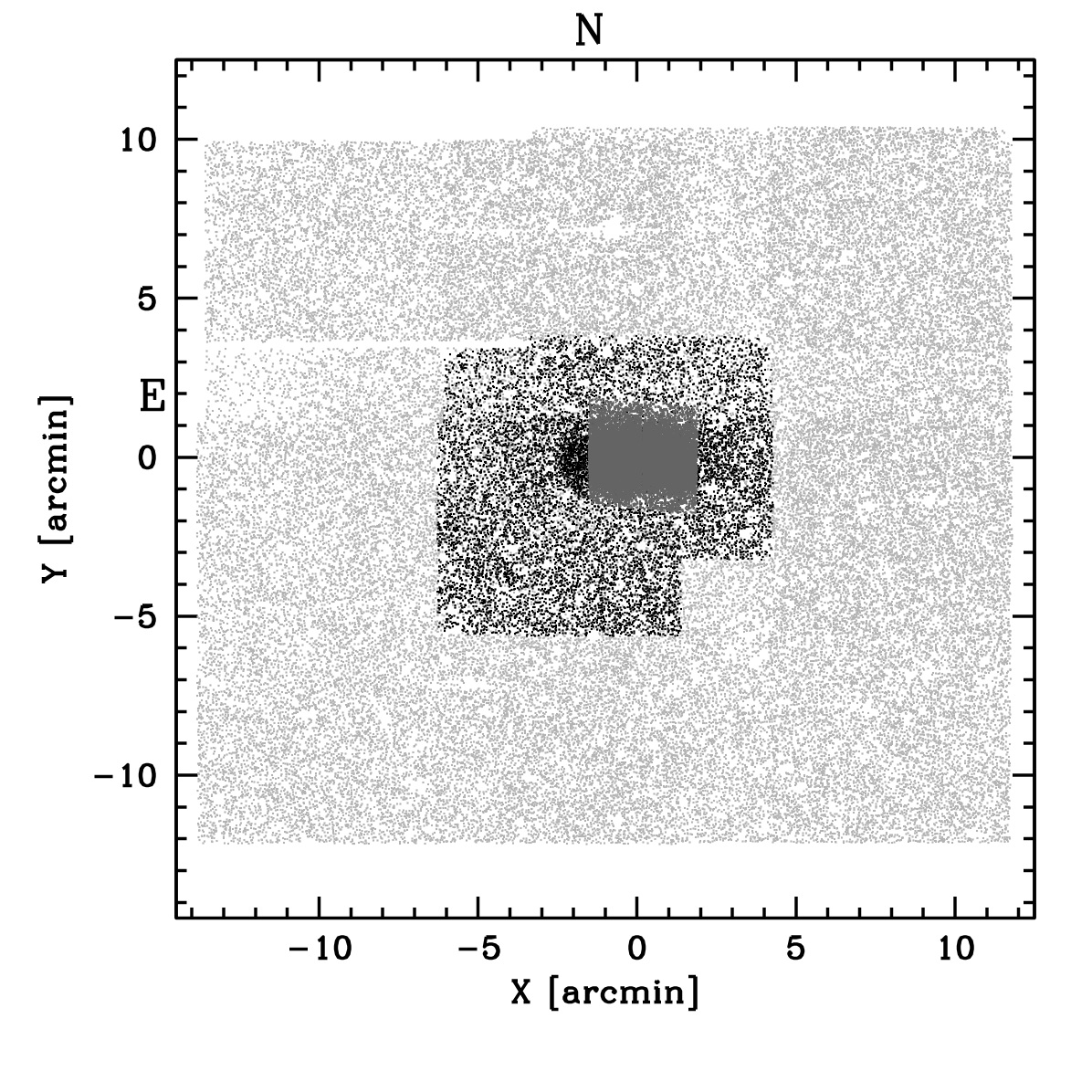}
     \caption{Maps of the stars in the various samples. Light grey: MOSAIC sample; black: IN6 sample;
     dark grey: HST-ACS sample }
        \label{mappet}
    \end{figure}

\begin{figure}
   \centering
   \includegraphics[width=\columnwidth]{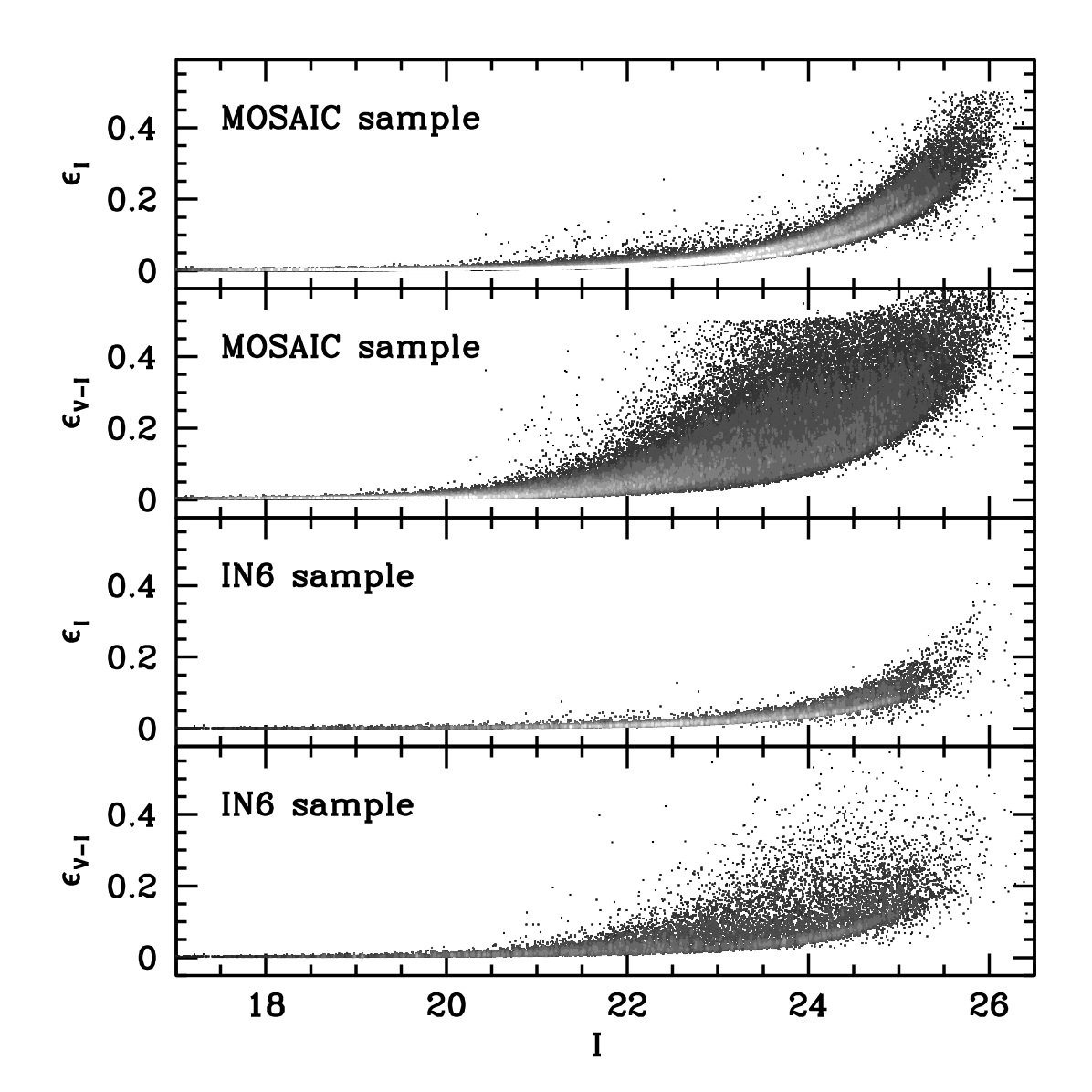}
     \caption{Distribution of photometric errors in magnitude and colours for the two samples,
     MOSAIC (upper two panels), and IN6 (lower two panels).Stars are plotted as black points in regions of the diagrams with a few stars and as grey squares otherwise, with the scale of grey proportional to the local density. Lighter tones of grey correspond to higher density.}
        \label{err}
    \end{figure}

In this case DAOPHOT/ALLFRAME was forced to produce an output catalog containing only sources  with at least {\em six} independent magnitude estimates per filter. This strong requirement limits the global FoV covered by the IN6 sample to an inner region that maximises the overlapping between chips from different pointings 
(see Fig.~\ref{mappet}). The average and the standard error of the mean of the independent measures obtained from the different images were adopted as the final values of the instrumental magnitude and of the uncertainty on the relative photometry. The large number of independent magnitude estimates combined imply a significantly higher precision of the relative photometry with respect to the MOSAIC sample, as is evident from Fig.~\ref{err}. Also in this case the instrumental coordinates were transformed into 
J2000 Equatorial coordinates with second degree polynomials, using more than 1000 stars in common with the GSC2.2 catalogue. The r.m.s of the fits is $\le 0.23\arcsec$ in each coordinate. We use the IN6 catalogue in Sect.~\ref{grad} to study the metallicity and populations gradients in Sgr~dIrr, and in Sect.~\ref{struc} to validate the structural analysis performed with the MOSAIC sample. We make the IN6 catalog publicly available at the CDS.

\subsection{Photometric calibration}

The instrumental magnitudes (v, i) of both samples were transformed into the Johnson-Kron-Cousins V,I  photometric system \citep{landolt} using more than 70 bright stars ($V<22.5$) in common with the publicly available HST Wide Field and Planetary Camera 2 (WFPC2) photometry by \citet{holtz}\footnote{\tt \tiny http://astronomy.nmsu.edu/holtz/archival/html/lg.html}, with first-order polynomials as a function of v-i for the I band, and with a simple zero point (ZP) correction for the V band. The transformations were optimised for the colour range spanned by stars of the target galaxy ($-0.5\la V-I\la 2.0$). 

We compared our final calibrated photometry with that by \citet[][LK00, hereafter]{lk00}, {\bf who provide} B,V,R,I magnitudes for a subsample of stars brighter than V=20.8. The mean difference between the two photometries is less than 0.01 in V from 85 stars in common, while it is larger than 0.2 in I with a significant dependence on colour. The comparison with the HST-ACS photometry presented below strongly suggests that our I photometry is more correct than it is in LK00. 

\subsection{HST-ACS photometry}

From the HST archive we retrieved the F606W and F814W images of Sgr~dIrr from the program GO~9820, i.e. part of the observational material used by M05. PSF photometry on the pre-reduced images was performed with the ACS module of DOLPHOT \citep{dolph}, exactly in the same way as described in detail in \citet{vvhst}.
As in that analysis, we only selected stars with the best quality flag ({\tt obj\_type=1}), $CHI<2.0$, $|SHARP|<0.5$, crowding parameter $<0.3$, and photometric error $<0.3$~mag, in both filters. These cuts produced a cleaned catalogue with 24911 sources. DOLPHOT automatically provides calibrated magnitudes, also in the Johnson-Kron-Cousins system that we adopt here. The final CMD is very similar to that obtained by M05 (see Fig.~\ref{HST}, below). The instrumental coordinates were transformed into 
J2000 equatorial coordinates with third degree polynomial, using more than 3300 stars in common with the IN6 catalogue. The r.m.s of the fits is $\le 0.05\arcsec$ in each coordinate. The position of the ACS sample with respect to the two VIMOS samples is illustrated in Fig.~\ref{mappet}.

\subsection{Reddening and distance}

We interpolated the \citet{ebv} reddening maps - as re-calibrated by \citet{schlaf} - to obtain an estimate of $E(B-V)$ and its variation over the MOSAIC FoV. We adopted a regular grid with knots spaced by $2.0\arcmin$, finding (a) a mean reddening values slightly lower than those usually reported in the literature (see, e.g., M05, M12) due to the Schlafy et al. re-calibration, and (b) that the reddening variations are negligible for our purposes (see Table~\ref{Tab_par}; see M05 for a discussion of the effect of differential reddening on the youngest stars). In the following, we adopt the reddening laws $A_V=3.1E(B-V)$ and $A_I=1.76E(B-V)$ from \citet{dean}. 

To obtain distances fully consistent with the scale adopted in Papers~I and II we used 
our newly derived V,I ACS-HST photometry to (a) correct the CMD for the newly derived values of the reddening, and (b) derive the magnitude and mean colours of the RGB-Tip again, as done in Papers~I and II. We obtain $I_{0,tip}= 21.10 \pm 0.04$ and $(V-I)_{0,tip}=1.27$, in excellent agreement with M05 (taking also into account  the slight difference in the adopted reddening). Using the relation for the absolute magnitude of the RGB Tip already adopted in the previous papers of this series:

\begin{equation}
M_I^{TRGB} = 0.080(V-I)_0^2 -0.194(V-I)_0 -3.939
\end{equation}

\noindent
from \citet{cefatip}, we obtain $M_I^{TRGB} = -4.06\pm 0.10$. Finally, we obtain $(m-M)_0 = 25.16 \pm 0.11$, corresponding to $D=1076\pm 55$~kpc, where the reported errors include {\em all the statistical and systematic sources of error}. The derived distance estimate is in excellent agreement with M05 and M12.

\subsection{Surface photometry and coordinate system}
\label{sbp}

As a basis for constructing the final composite  profiles 
(surface photometry + star counts; see Sect.~\ref{struc}), we adopted the surface photometry by LK00. 
We also adopted the ellipticity ($\epsilon$) and  position angle (PA) estimated by LK00, since we verified that they are appropriate over the whole radial range where we detect Sgr~dIrr stars (see Sect.~\ref{struc} below).

Adopting the coordinates for the centre listed in Table~\ref{Tab_par}, we
converted the equatorial coordinates of each star ($\alpha$,$\delta$) to cartesian coordinates $X$,$Y$(in arcmin) on the plane of sky as 
done in Paper~I, with $X$ increasing towards the west and $Y$ increasing towards the north. 
Finally, we defined the elliptical distance from the centre of the galaxy (or elliptical radius $r_{\epsilon}$) as

\begin{displaymath}
r_{\epsilon}=\sqrt{X^2+{\Big(\frac{Y}{1-\epsilon}\Big)}^2}
\end{displaymath}

\noindent
which is equivalent to the major-axis radius. The $X$,$Y$ coordinate system and $r_{\epsilon}$ will always be adopted in the following analysis.

\begin{table}
  \begin{center}
  \caption{Observed and derived parameters of Sagittarius dIrr}
  \label{Tab_par}
  \begin{tabular}{lcr}
    \hline
    Parameter & Value & Notes\\
\hline
$\alpha_0$     & 19:29:59.9                     & J2000$^a$  \\
$\delta_0$     & -17:40:49.4                    & J2000$^a$ \\
$l_0$          & $21.055\degr$                  & Gal. long.\\
$b_0$          & $-16.289\degr$                   & Gal. lat.\\
SGL            & $221.275\degr$                   & Supergal. long.\\
SGB            & $55.516\degr$                  & Supergal. lat.\\
E(B-V)         & $0.107\pm 0.010$               & Average $\pm \sigma^b$ \\
$(m-M)_0$      & $25.16\pm 0.11$                &   \\
D              & $1076\pm 55$ kpc             &   \\
1~arcsec       & 5.2~pc                         & conv. factor at D=1076~kpc\\
$V_{\rm h}$    & $-78.5\pm 1$~km~s$^{-2}$         & heliocentric velocity$^c$ \\
$V_{g}$        & $8$~km~s$^{-1}$              & galactocentric velocity$^c$ \\
$\langle{\rm [Fe/H]}\rangle$ & $-2.0 $       & from RGB colour$^d$ \\
$\epsilon$     & $0.50$                         & from LK00  \\
PA             & $90\degr$                      & from LK00 \\
$\mu_V(0)$     & $ 23.9\pm 0.1$ mag/arcsec$^2$ & central SB$^e$   \\ 
$h$          & $1.1\arcmin$                  & S\'ersic scale radius \\
$r_h$        & $1.16 \arcmin$                   & Observed$^f$  \\
$V_{tot}$      & $13.4 \pm 0.1$                &  Observed$^f$ \\
$M_V$          & $-12.1 \pm 0.2$               &   \\
$L_V$          & $5.8^{+1.1}_{-1.0}\times 10^6~L_{V,\sun}$    & total V luminosity \\
$M_{HI}$       & $8.7\times 10^6~M_{\sun}$      & \HI~ mass$^g$  \\
\hline
\end{tabular}
\tablefoot{$^a$ Estimated by eye from HST-ACS images and verified to provide a more
satisfactory centre of symmetry for the distribution of RGB stars w.r.t. the centre listed by M12.
Our new centre is $13.5\arcsec$ to the east and $9.4\arcsec$ to the south of the M12 centre. 
$^b$ Average over the whole FoV from the reddening maps by \citet{ebv}, as re-calibrated by 
\citet{schlaf} (regular grid covering the whole MOSAIC field with knots spaced by $2\arcmin$). 
$^c$ From \citet{mcc}.
$^d$ From the HST-ACS sample, see sect.~\ref{grad}. 
$^e$ $\mu_V$ of the innermost point of the observed profile. Not corrected for extinction. 
The central SB of the best fitting Sersic profile is $\mu_V(0)=23.3$mag/arcsec$^2$. 
$^f$ From the numerical integration of the SB profile in $r_{\epsilon}$. $V_{tot}$ is in excellent agreement with the integrated magnitude of the best-fit Sersic model, $V_{tot}=13.3$. 
$^g$ From \citet{littlethings}.} 
\end{center}
\end{table}

\subsection{The color-magnitude diagrams}

\begin{figure}
   \centering
   \includegraphics[width=\columnwidth]{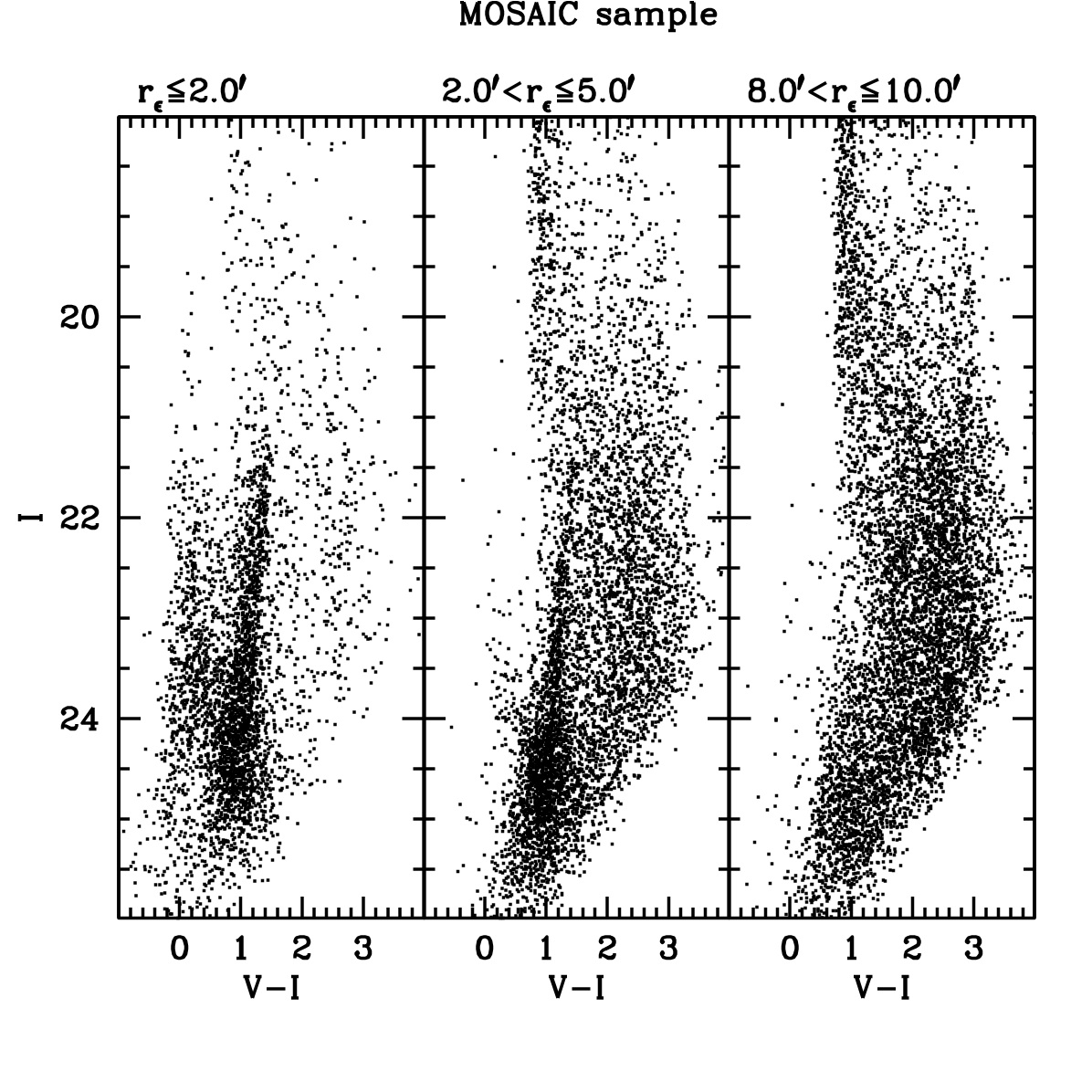}
     \caption{CMDs in different radial ranges from the MOSAIC sample.}
        \label{cmds}
    \end{figure}

In Fig.~\ref{cmds} we present our final CMD in different radial ranges for the MOSAIC sample. Our photometry reaches $I\sim 25.5$, thus sampling the extended red clump (RC) of Sgr~dIrr.
The CMDs is dominated by a strong and (relatively) wide red giant branch (RGB), ranging from (I, V-I)$\sim(25.5,0.8)$ to $\sim(21.5,1.5)$. The population of He-burning stars extends from the RC up to (I, V-I)$\sim(19.5,0.0)$ into a blue loop (BL) sequence. The young main cequence (MS) plume at the blue edge of the CMDs 
runs parallel to the BL sequence (and partially overlapping it, in colour) up to $I\sim 21$. The sparse sequence of red super giants (RSG) is barely visible, on the blue side of the RGB and parallel to it for
$I\la 22.5$ (see M05 for a detailed description of the various evolutionary sequences that can be identified in the ACS CMD).

The middle panel of Fig.~\ref{cmds} clearly demonstrates that Sgr~dIrr stars are present in a (relatively) large number beyond the radial range covered by existing surface photometry ($r_{\epsilon}\simeq 2\arcmin$,
where $\mu_V\simeq 26.3$~mag/arcsec$^2$, see LK00 and Fig.~\ref{prof}, below). The right-hand panel of the figure shows the  fore/back-ground population that contaminates our CMDs. 
Given the relatively low angular distance both from the Galactic centre and plane, the contamination from foreground Galactic stars is particularly severe, e.g., with respect to the high latitude galaxies studied in Papers~I and II.
The vertical plume spanning the whole diagram around V-I$\sim 3.0$ is due to local M dwarfs; the sparse wide band around V-I$\simeq 1.0$, bending to the red for $I\ga 22.0$, is due to foreground MS stars in the Galactic thick disc and halo; while the broad blob of sources with $0.0\la$ V-I$\la 2.0$ and $I\ga 23.0$ is largely dominated by distant unresolved  galaxies (see Paper~I for a detailed discussion and references).

\begin{figure}
   \centering
   \includegraphics[width=\columnwidth]{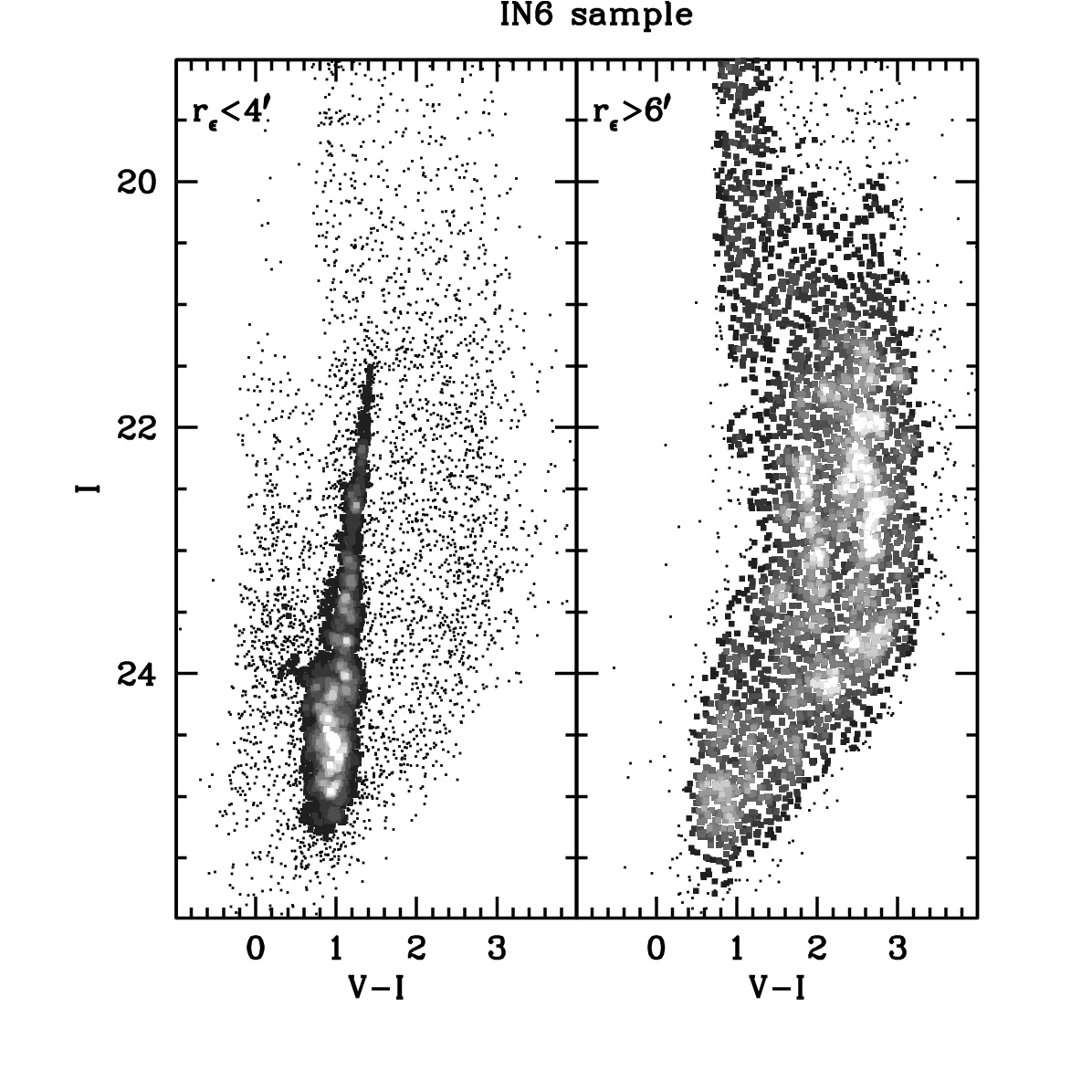}
     \caption{CMDs in different radial ranges from the IN6 sample.
     Stars are plotted as black points in regions of the CMD with a few stars and as grey squares, otherwise with the scale of grey proportional to the local density in the CMD. Lighter tones of grey correspond to higher density.}
        \label{cmdIN6}
    \end{figure}

In Fig.~\ref{cmdIN6} the CMDs for two large radial ranges of the IN6 sample is presented as a greyscale
Hess diagram \citep[as in][]{vvhst}. Some features are more clearly seen in this diagram, such as the density peak of the RC and a weak over-density above the RGB Tip that may trace the AGB population identified by M05 and \citet{gulli}.

\begin{figure}
   \centering
   \includegraphics[width=\columnwidth]{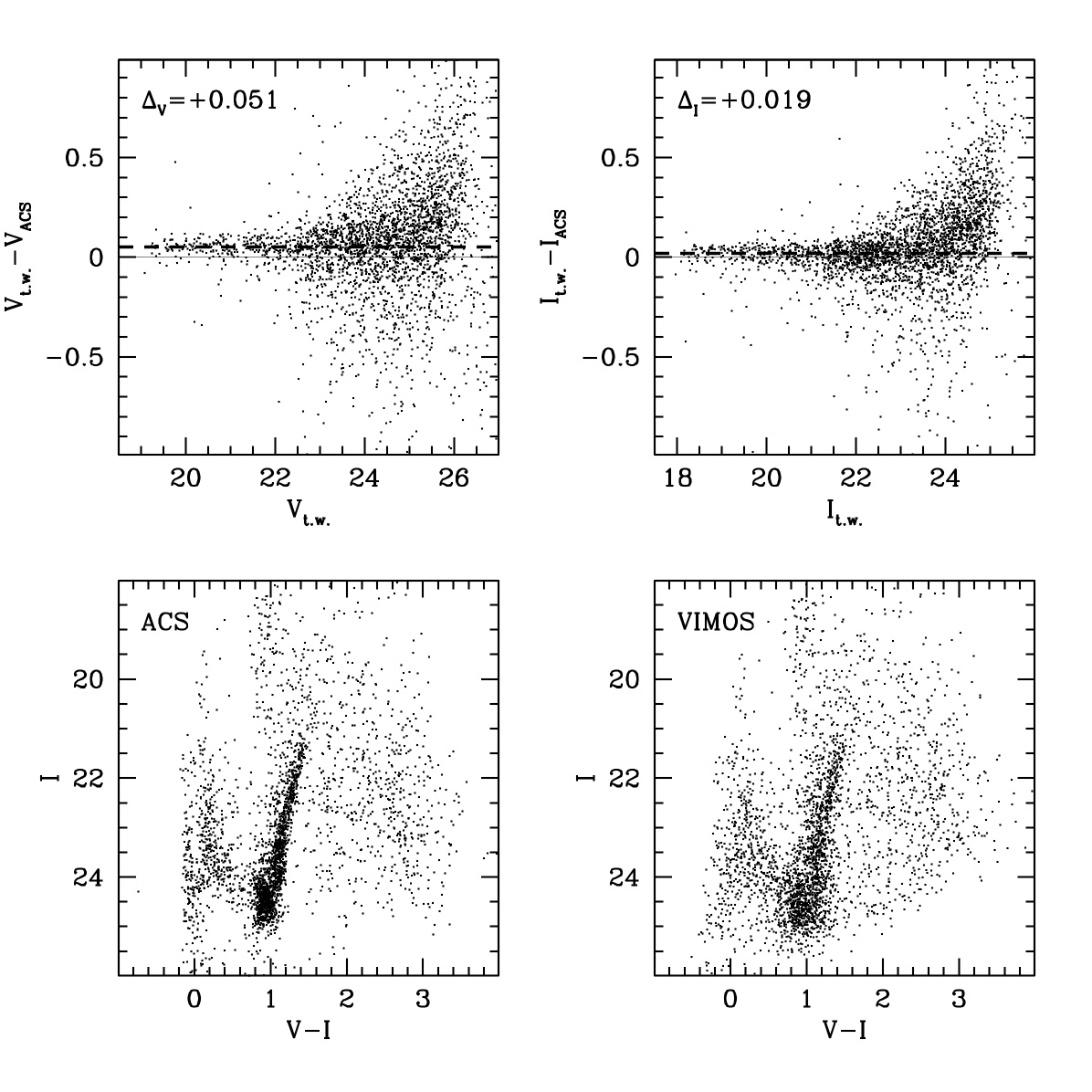}
     \caption{Comparison between HST-ACS and VLT-VIMOS photometry. Upper panels: comparison of the photometric zero points. Lower panels: CMDs for the stars in common between the two datasets.}
        \label{HST}
    \end{figure}

In Fig.~\ref{HST} we provide a direct comparison between VIMOS (IN6) and HST-ACS photometry. In the upper panels, it is shown that the photometric ZPs are in reasonable agreement. 
The lower panels display the CMDs obtained from the two datasets for the stars in common. Clearly our ground based photometry cannot rival the top-level quality achievable from the space, especially in the highly crowded central regions sampled by the HST data.
Still, it is remarkable that virtually all the features that can be identified in the VIMOS CMDs  have clear counterparts in the HST CMDs, thus indicating that the precision of
our photometry is sufficient to reliably distinguish stars in different evolutionary phases. 

\section{Metallicity and population gradients}
\label{grad}

\begin{figure}
   \centering
   \includegraphics[width=\columnwidth]{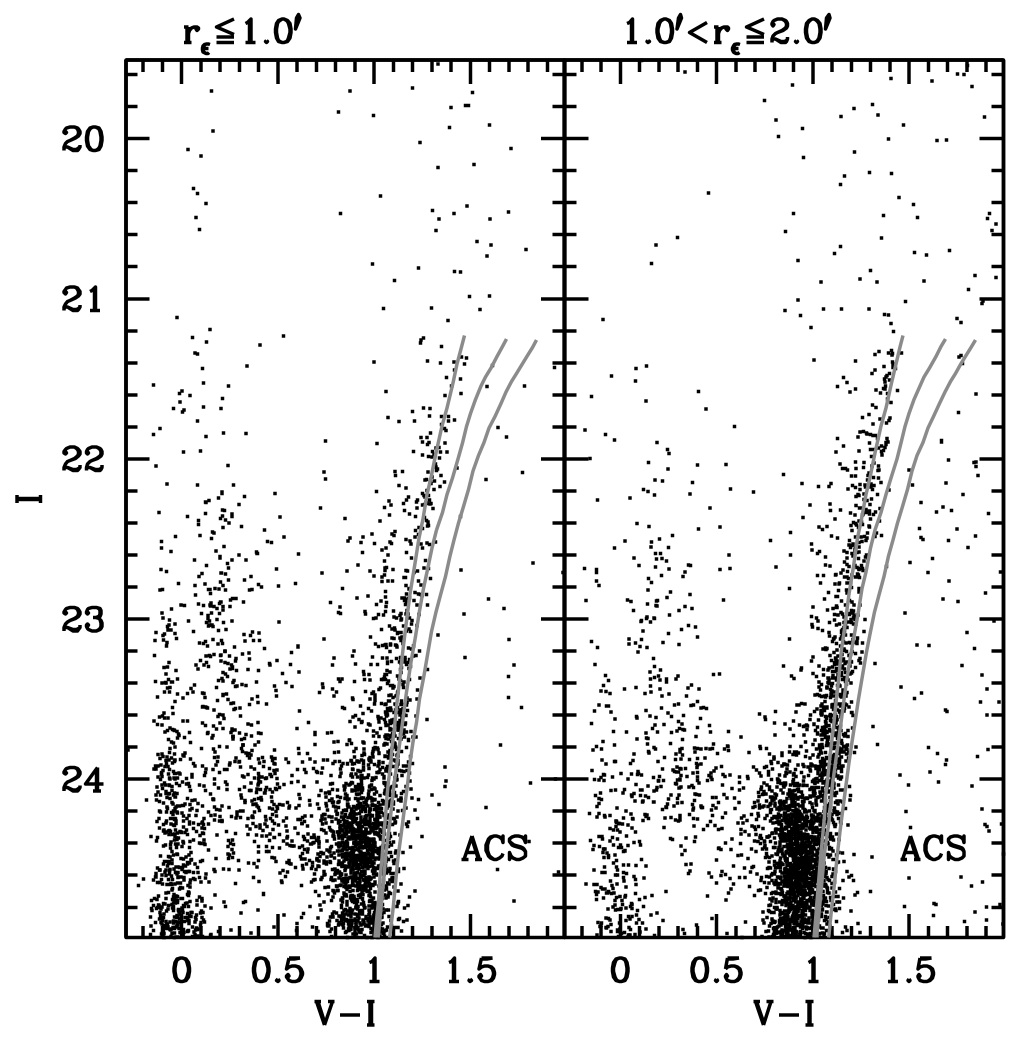}
     \caption{CMDs zoomed on the RGB for two radial ranges, from the HST-ACS sample. The ridge lines of the clusters NGC6341, NGC5275, and NGC288, from left to right, from the set by \citet{saviane} have been superimposed, after shifting them to reddening and distance of Sgr~dIrr. According to the December 2010 version of the \citet{harris} catalogue, the metallicity values corresponding to the three ridge lines are [Fe/H]=-2.31, -1.50, and -1.32, respectively.}
        \label{ACSmet}
    \end{figure}

In Fig.~\ref{ACSmet} we compare the observed RGB from the HST-ACS sample in two radial ranges covered by these data with a grid of RGB ridge lines of Galactic globular clusters (GC). The ridge lines are taken from the homogeneous set by \citet{saviane}; the adopted reddening and distance of the clusters are from \citet{f99}.
The template clusters are NGC6341, NGC5275, and NGC288, from left to right; the associated metallicities are   
[Fe/H]=-2.31, -1.50, and -1.32, respectively, from the December 2010 version of the \citet{harris} catalogue.
In both CMDs the [Fe/H]=-2.31 template lies straight in the middle of the bulk of RGB stars, and there is no hint of a difference in the RGB colours in the considered radial ranges. Interpolating on this grid of RGB templates \citep[as done in Papers~I and II, following][]{umi} for RGB stars within two magnitudes from the tip (and neglecting the effects of field contamination, which should be very weak for this sample), we obtain a mean metallicity $\langle[Fe/H]\rangle = -2.0$ with a standard deviation $\sigma_{[Fe/H]}=0.37$, in good agreement with M05 \citep[see also][and references therein]{moma2}. While there is no doubt that the  galaxy is dominated by very metal-poor stars 
\citep[see, e.g.][and references therein, for the metallicity of HII regions]{saviH2}, the derived mean should be regarded as a lower limit to the mean metallicity, since the observed RGB is also significantly populated by stars that are a few Gyr younger than the GC templates (see Paper~II for a detailed discussion of the analogous case of Sex~A). The presence of intermediate-age populations in Sgr-dIrr is demonstrated by the morphology of the red clump and by the presence of bright AGBs and carbon stars \citep[see][and references therein]{gulli}. Photometric estimates of the mean metallicity in these distant stellar systems with extended star formation should always be considered as educated guesses.

   \begin{figure}
   \centering
   \includegraphics[width=\columnwidth]{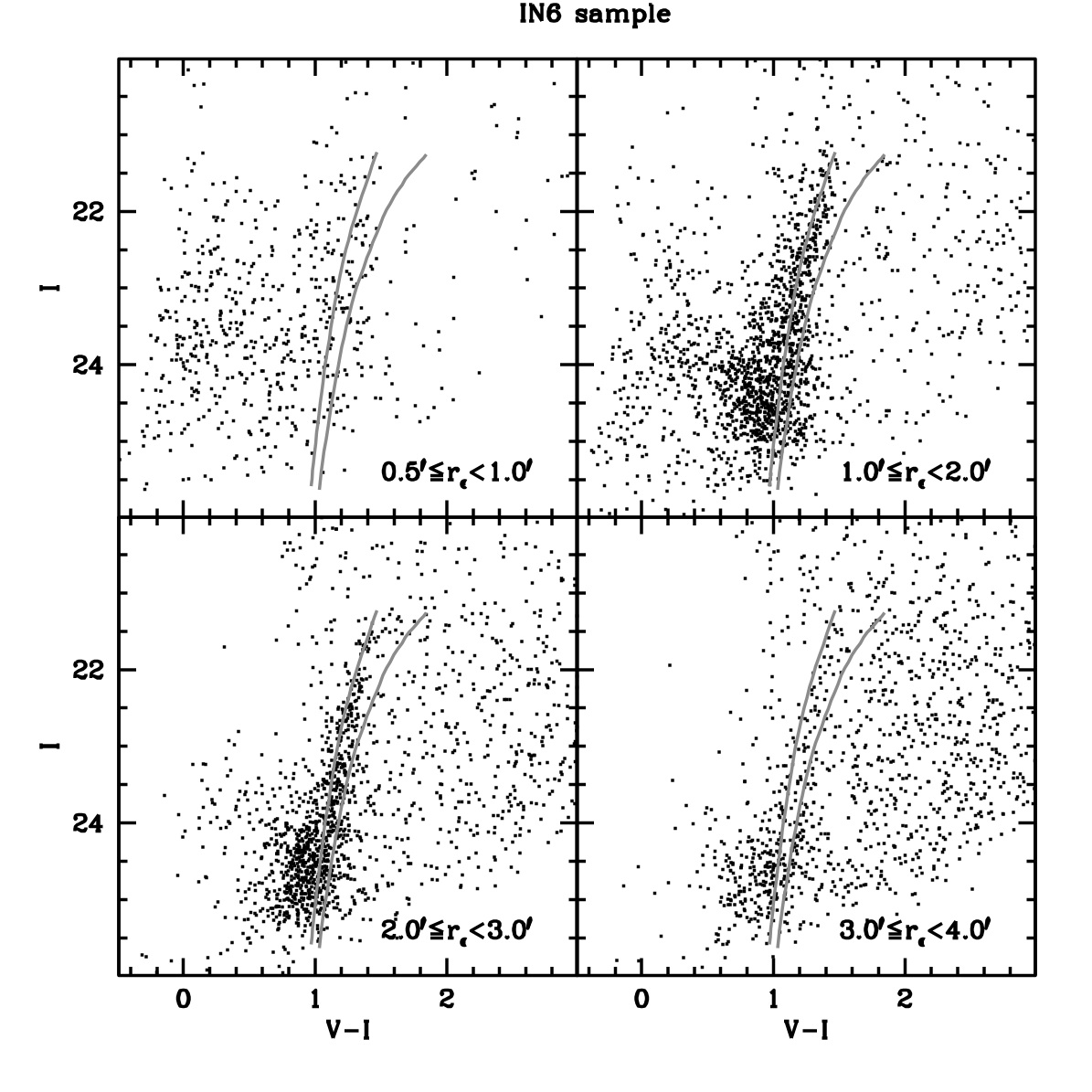}
     \caption{CMDs zoomed on the RGB for four radial ranges, from the VIMOS IN6 sample. The ridge lines of the clusters NGC6341 and NGC288, from left to right, from the set by \citet{saviane} have been superimposed, after shifting them to reddening and distance of Sgr~dIrr.}
        \label{cmmet}
    \end{figure}


We extend the search for a metallicity gradient to larger elliptical radii using the IN6 sample. In Fig.~\ref{cmmet}
we compare the CMDs for four different radial ranges, reaching the outer limit of Sgr~dIrr ($r_{\epsilon}=4.0\arcmin$, see Sect.~\ref{struc}, below) with the same grid of RGB templates used in Fig.~\ref{ACSmet}.
We removed the ridge line of NGC5272, for clarity. In the innermost radial range, the effects of crowding 
are too severe to allow a meaningful comparison. On the other hand, the results from the HST-ACS and IN6 samples for the range $1.0\arcmin<r_{\epsilon}\le 2.0\arcmin$ are fully compatible.
From Figs.~\ref{ACSmet} and \ref{cmmet} we can conclude that we do not detect any sign of a metallicity gradient in the RGB population over the whole body of the Sgr~dIrr galaxy.

   \begin{figure}
   \centering
   \includegraphics[width=\columnwidth]{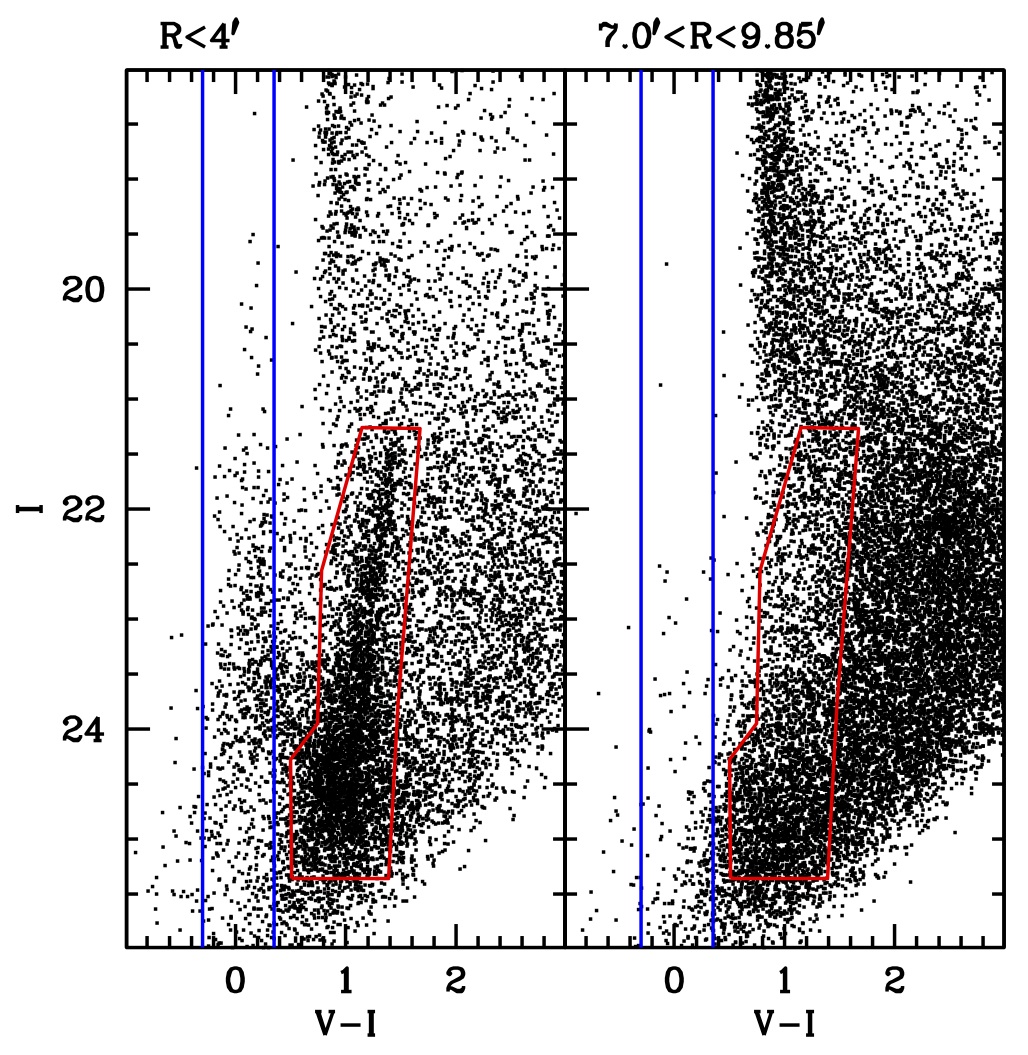}
   \includegraphics[width=\columnwidth]{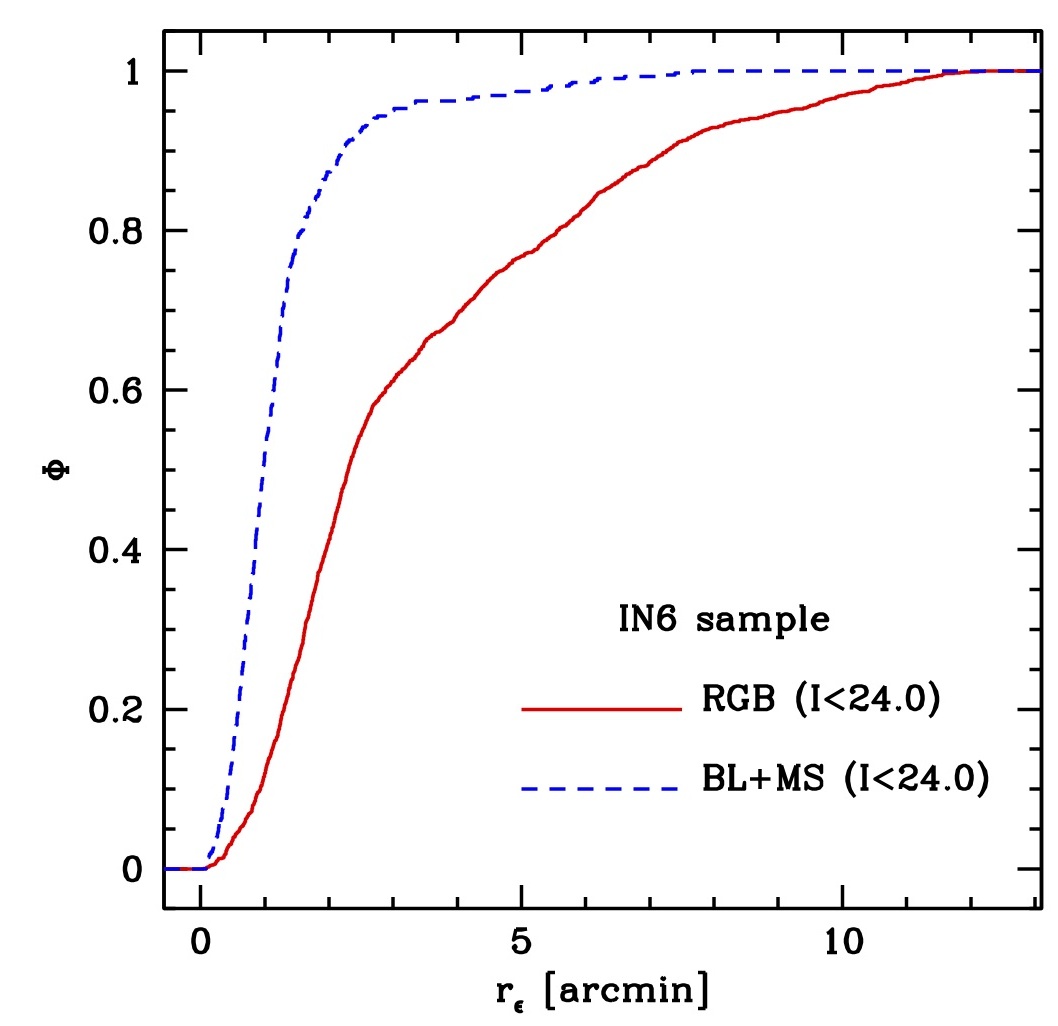}
     \caption{Upper panel: Selection boxes for BL+MS stars and RGB stars are superimposed to CMDs representative of
     the Sgr~dIrr galaxy (left panel) and of the fore/background contaminating field (right panel), from the MOSAIC sample. 
     The color range adopted to select BL+MS stars is $-0.30<V-I<0.35$.
     Note that the {\em field} CMD samples an area that is exactly three time of that sampled by the {\em on target} CMD. Lower panel: Cumulative radial distribution of the BL+MS (dashed blue line) and RGB (continuous red line) populations in the same range of magnitude.}
        \label{raddist}
    \end{figure}


In the lower panel of Fig.~\ref{raddist}, we compare the radial distribution of RGB and MS+BL stars, selected with the boxes illustrated in the upper panel of the same figure. The analysis is limited to stars in the same range of magnitude, for stars $\sim 2$~mag above the limits of the photometry, to compare samples that  are equally affected by incompleteness. As is typical of dwarf galaxies, the backbone of old-to-intermediate-age stars has a much more extended distribution than populations younger than 1~Gyr, as traced by MS+BL stars. Figure~\ref{raddist} shows that $\sim 90\%$ of MS+BL stars are confined within $r_{\epsilon}=2.0\arcmin$, a region that is largely covered by the HST-ACS sample. For a detailed analysis of this population and of its spatial distribution in relation with the distribution of \HI, ~we refer the interested reader to the extensive discussion by M05.

Finally, M05 notes a gap in luminosity in the brightest part of the BL sequence (between I=21.1 and I=20.6 in Fig.~\ref{ACSmet}, which may trace a short quiescent phase in the star formation that occurred some 2--60~myr ago. Unfortunately, the population on the bright side of the gap is constituted by a handful of stars, so it is difficult to firmly assess the statistical significance of this feature. In this context, it is interesting to note that the same gap is clearly detected in our own ground-based photometry, in particular when the IN6 sample is considered (see Fig.~\ref{cmdIN6}), thus providing additional support to a scenario of intermittent star formation in the past 100~Myr, which might also be associated to the inhomogeneities in the distribution and velocity field of the \HI~(see Sect.~\ref{HI}).

\section{Structure}
\label{struc}

\begin{figure}
   \centering
   \includegraphics[width=\columnwidth]{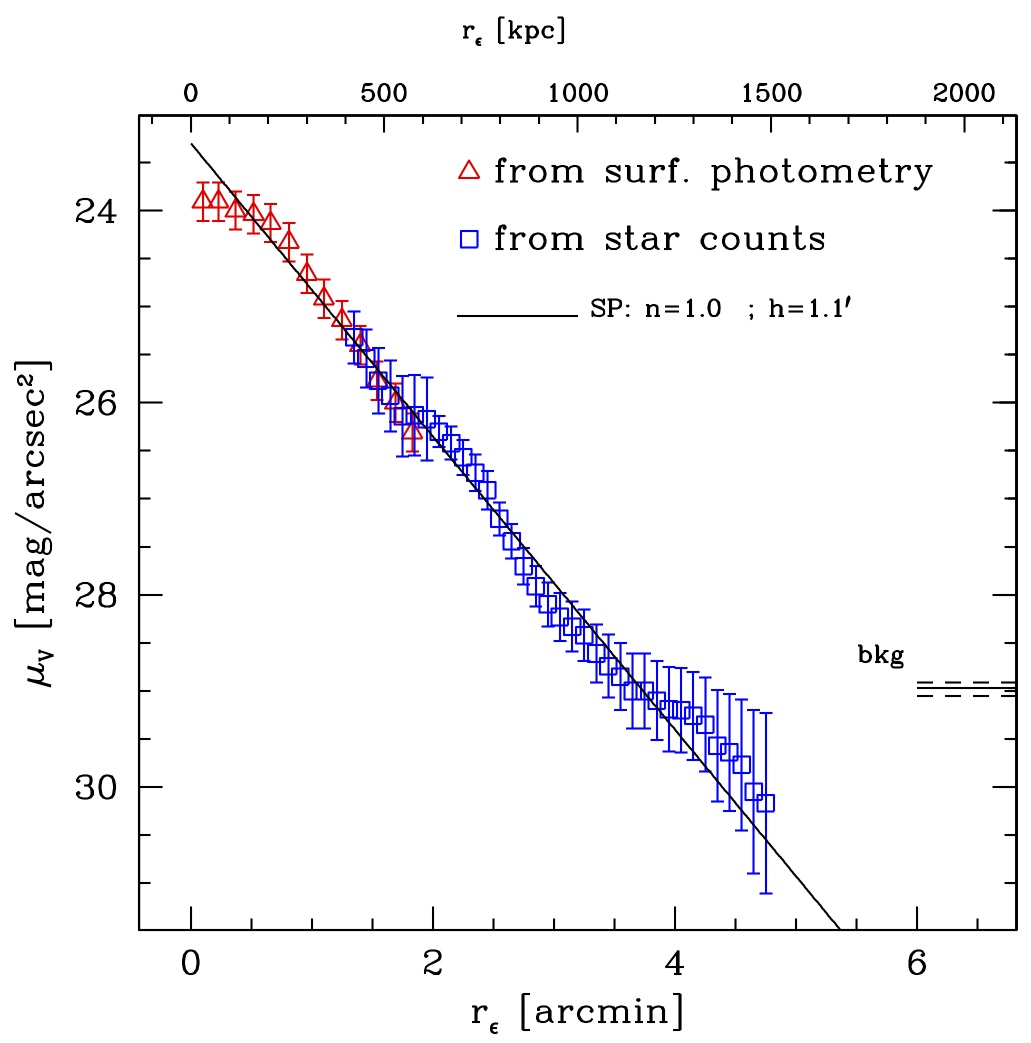}
     \caption{Surface brightness profile of Sgr~dIrr  in V band 
 obtained by joining surface photometry from LK00 (open triangles) and star count (open squares) profiles. The continuous line is the Sersic model that best fits the profile.  The level of the background (bkg) and the associated uncertainty is also reported.}
        \label{prof}
    \end{figure}

In the lower panel of Fig.~\ref{prof} we show the azimuthally-averaged surface brightness (SB) profile of Sgr~dIrr. The profile was obtained  by joining the V-band surface photometry from LK00  out to $r_{\epsilon}= 1.83\arcmin$, with the surface density profile obtained from star counts  in elliptical annuli, while keeping the values of $\epsilon$ and $PA$ derived in Sect.~\ref{sbp} fixed. The (large) overlapping region between the profiles from surface photometry and from star counts  was used to normalise the star-count profile, shifting it to the V magnitude scale of the surface photometry profile. 

For the star counts we used as homogeneous density tracers, candidate RGB stars enclosed in the polygonal box which is shown in the upper panel of Fig.~\ref{raddist}. The background was estimated in wide areas at $r_{\epsilon}>8\arcmin$. The profiles were constructed starting with bright magnitude limits in the inner regions and going to fainter limits in the outskirts to get rid of variations in the completeness with distance from the galaxy centre and to have the highest SB sensitivity in the low-SB external regions, as described in detail in Paper~II.

As in the cases presented in previous papers of this series, our newly derived composite profile reaches much larger radii (by a factor of $\ga 2$) than previous studies, which were based solely on surface photometry.
In particular, we are able to trace the stellar profiles out to $r_{\epsilon}\simeq 4.7\arcmin$, showing that this galaxy is also much more extended than what previously observed (see Papers~I and II, and references therein).

\begin{table}
  \begin{center}
  \caption{Observed surface brightness profile.}
  \label{sbprof}
  \begin{tabular}{lcc}
$r_{\epsilon}$ & $\mu_V$ &Source       \\
arcmin         & mag/arcsec$^2$  \\
  0.10 &  23.91 $\pm$  0.10 & LK00  \\
  0.23 &  23.91 $\pm$  0.10 & LK00  \\
  0.37 &  24.00 $\pm$  0.10 & LK00  \\
  0.52 &  24.04 $\pm$  0.10 & LK00  \\
  0.66 &  24.13 $\pm$  0.10 & LK00  \\
  0.81 &  24.33 $\pm$  0.10 & LK00  \\
  0.96 &  24.66 $\pm$  0.10 & LK00  \\
  1.10 &  24.92 $\pm$  0.10 & LK00  \\
  1.25 &  25.14 $\pm$  0.10 & LK00  \\
  1.40 &  25.40 $\pm$  0.10 & LK00  \\
  1.54 &  25.77 $\pm$  0.10 & LK00  \\
  1.69 &  26.00 $\pm$  0.10 & LK00  \\
  1.83 &  26.31 $\pm$  0.10 & LK00  \\
  1.95 &  26.17 $\pm$  0.43 &  sc   \\
  2.05 &  26.30 $\pm$  0.16 &  sc   \\
  2.15 &  26.42 $\pm$  0.17 &  sc   \\
  2.25 &  26.57 $\pm$  0.18 &  sc   \\
  2.35 &  26.73 $\pm$  0.19 &  sc   \\
  2.45 &  26.91 $\pm$  0.20 &  sc   \\
  2.55 &  27.21 $\pm$  0.17 &  sc   \\
  2.65 &  27.44 $\pm$  0.18 &  sc   \\
  2.75 &  27.70 $\pm$  0.19 &  sc   \\
  2.85 &  27.91 $\pm$  0.21 &  sc   \\
  2.95 &  28.10 $\pm$  0.23 &  sc   \\
  3.05 &  28.23 $\pm$  0.25 &  sc   \\
  3.15 &  28.33 $\pm$  0.26 &  sc   \\
  3.25 &  28.42 $\pm$  0.27 &  sc   \\
  3.35 &  28.61 $\pm$  0.30 &  sc   \\
  3.45 &  28.74 $\pm$  0.33 &  sc   \\
  3.55 &  28.85 $\pm$  0.35 &  sc   \\
  3.65 &  29.00 $\pm$  0.39 &  sc   \\
  3.75 &  29.00 $\pm$  0.39 &  sc   \\
  3.85 &  29.10 $\pm$  0.41 &  sc   \\
  3.95 &  29.19 $\pm$  0.44 &  sc   \\
  4.05 &  29.20 $\pm$  0.44 &  sc   \\
  4.15 &  29.26 $\pm$  0.46 &  sc   \\
  4.25 &  29.35 $\pm$  0.49 &  sc   \\
  4.35 &  29.57 $\pm$  0.58 &  sc   \\
  4.45 &  29.64 $\pm$  0.61 &  sc   \\
  4.55 &  29.77 $\pm$  0.68 &  sc   \\
  4.65 &  30.05 $\pm$  0.85 &  sc   \\
  4.75 &  30.17 $\pm$  0.94 &  sc   \\
\hline
\end{tabular} 
\tablefoot{LK00 = surface photometry from LK00; sc = star counts from the 
present analysis. Since LK00 do not provide uncertainties, we assume 0.1 mag for 
their estimates. Not corrected for extinction.}
\end{center}
\end{table}

   \begin{figure}
   \centering
   \includegraphics[width=\columnwidth]{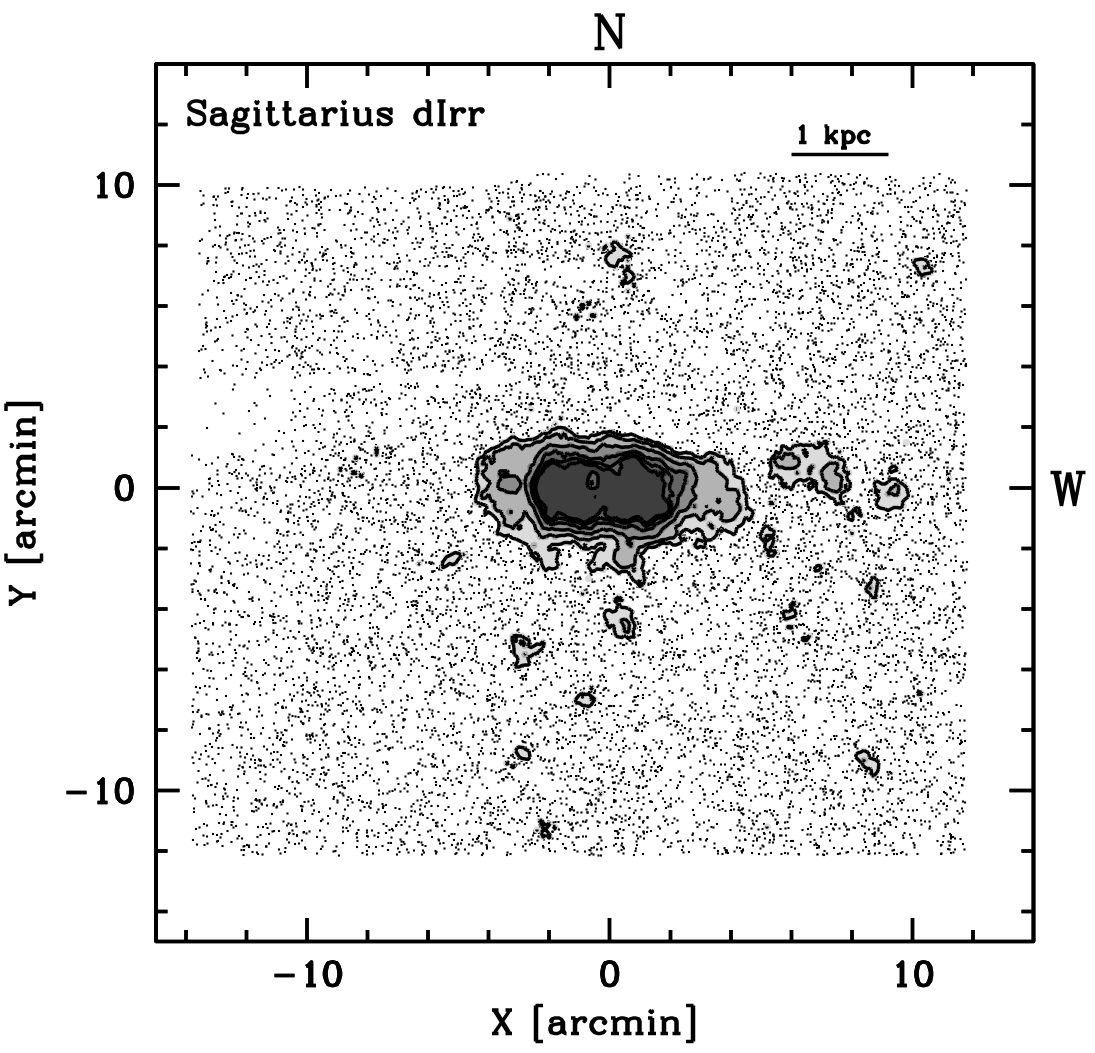}
     \caption{Density map obtained from RGB star counts. The levels of grey correspond to density of 3, 5, 10, 20, 40, and 80$\sigma$ above the background, from the lightest to the darkest tone of grey.  All the RGB stars falling in the selection box defined in Fig.~\ref{raddist} are also plotted (as dots) to provide a direct illustration of the effects of the footprint of the mosaic.
     }
        \label{maps}
    \end{figure}


In general, the overall profile is well fitted by a single \citet{sersic} model\footnote{Here we adopt the formalism by \citet{ciotti}, with the SB profile given by $I(r)=I_0e^{-b\eta^{\frac{1}{n}}}$ (their Eq.~1), with $\eta=\frac{r}{h}$, $h$ being the half-light radius. The value of the constant $b$ is from their Eq.~25, which works well for $n\ga 1.0$.} with $n=1.0$, i.e. a pure exponential, as is typical of low luminosity dwarfs \citep{mateo}.
For $r_{\epsilon}\ga 2\arcmin$ the profile from star counts present some weak kinks and weaves (with a hint of an external flattening of the slope) that we do not regard as significant, since small-scale asymmetries (both real and/or associated to holes due to heavily saturated foreground stars) can be the origin of these features in the low SB realm considered here. The flattening in the innermost $\sim 200$~pc is very similar to what is observed in the profile of Sextans~B (Paper~II).

Integrating the profile shown in Fig.~\ref{prof} numerically, we obtain integrated V magnitudes
$V_{tot}=13.4\pm 0.1$, and half-light radii $r_h=1.1\arcmin$, in excellent agreement with the respective values for the best-fitting exponential model and in reasonable agreement with the values reported in the literature (M12), which are based on much less extended observed profiles.

\subsection{Density maps}
\label{dmap}

In Fig.~\ref{maps} we present the density map for Sgr~dIrr obtained with the MOSAIC sample from RGB star counts (selected as for the SB profile) on a fixed regular grid of nodes spaced by $0.1\arcmin$, 
with the same adaptive algorithm described in Papers~I and II.
The lightest tone of grey corresponds to an over-density at $3\sigma$ above the background level, and it is the outermost density contour that we can reliably trace with our data. All the RGB stars
selected for the analysis are also plotted in the maps to give an idea of the
sampling biases that are obviously affecting them (inter-chip gaps, saturated
stars, etc.). Single isolated weak
peaks with typical scales $\la 1\arcmin$ can correspond to clusters of galaxies
whose red sequence galaxies fall in the RGB selection box or to clumping of spurious sources in the haloes of heavily saturated stars. We note, however, that we were unable to associate the wide blob centred at (X,Y)$\sim$(6.5,0.5) with any obvious clustering of galaxies in the deep stacked images. It therefore cannot be excluded that the stellar distribution of Sgr~dIrr may extend even beyond $X\sim 5\arcmin$, at surface brightness levels $\mu_V\ga 30.0$~mag/arcsec$^2$.

   \begin{figure}
   \centering
   \includegraphics[width=\columnwidth]{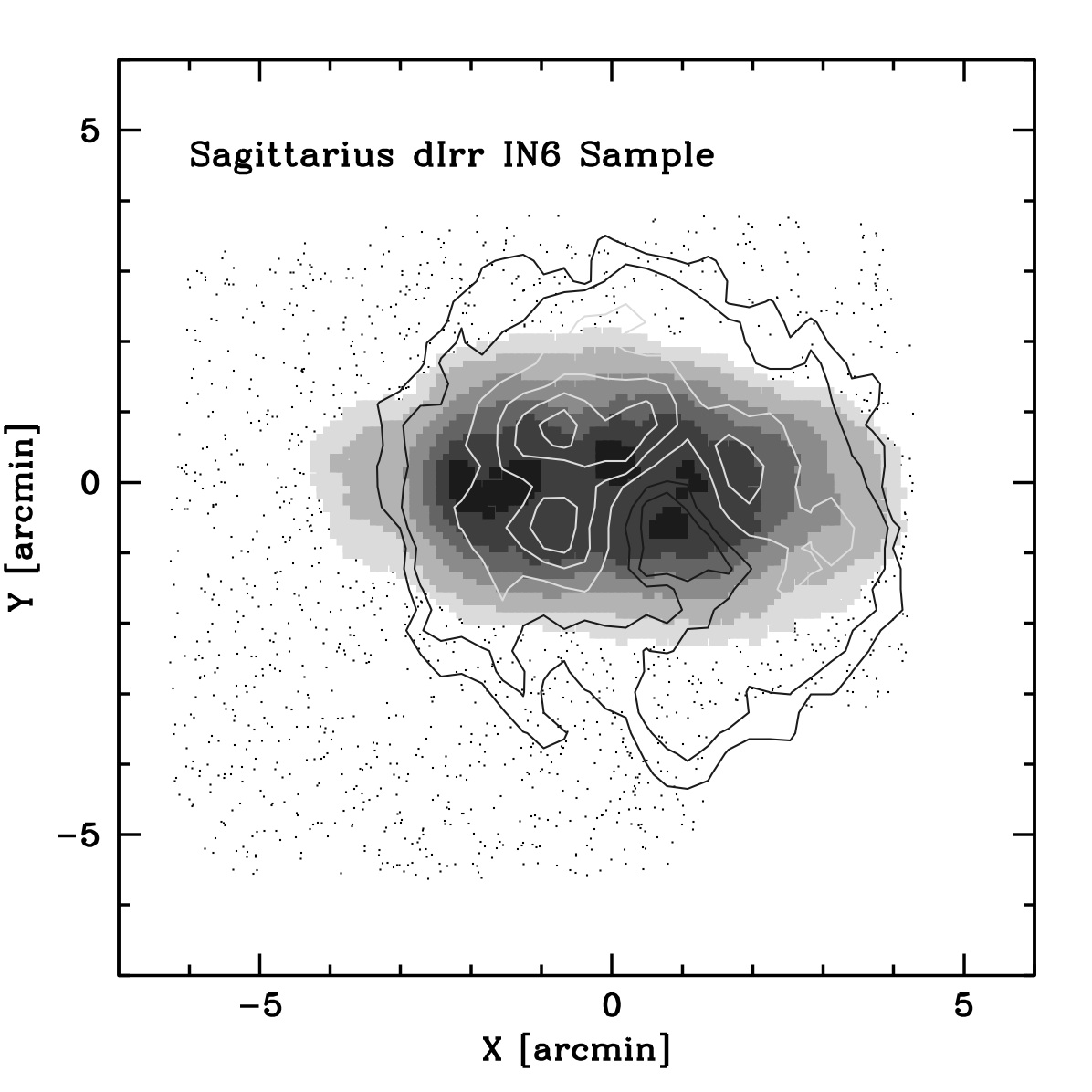}
     \caption{Density map obtained with the matched filter technique from the IN6 sample. The symbols are the same as in Fig.~\ref{maps}. A higher level of smoothing has been adopted. The \HI~ column density contours obtained from LITTLE THINGS data (with a beam of $28.2\arcsec\times 16.9\arcsec$) were superimposed. The associated density is 0.5, 1.0, 3.0, 5.0, 7.0, 9.0 $\times10^{20}$~cm$^{-2}$, from the outermost to the innermost contour.
     }
        \label{maps6}
    \end{figure}


The stellar body of Sgr~dIrr (as traced by intermediate-to-old stars) is made up of an inner boxy component, out to $r_{\epsilon}\simeq 2.0\arcmin$, surrounded by a more elongated halo, somehow resembling a thick disc seen nearly edge-on (see Paper~I).
This impression is confirmed by the map shown in Fig.~\ref{maps6} obtained with the matched filter technique
\citep[see][and Paper~I]{connie} from the IN6 sample, which is much less affected by discontinuities in the spatial sampling or by spurious sources than the MOSAIC sample.

It is interesting to note that the correlation with the distribution of \HI~ \citep[superimposed contours obtained from LITTLE THINGS data,][;see Sect.~\ref{HI}]{littlethings} is relatively poor. The \HI\ cloud associated to Sgr~dIrr has a size comparable to the stellar body, but its outer contours have a much rounder shape and irregular structure; the density peak is significantly off-centred with respect to the centre of symmetry of the stellar body.

\section{Neutral hydrogen}
\label{HI}

\begin{figure*}
   \centering
   \includegraphics[width=\columnwidth]{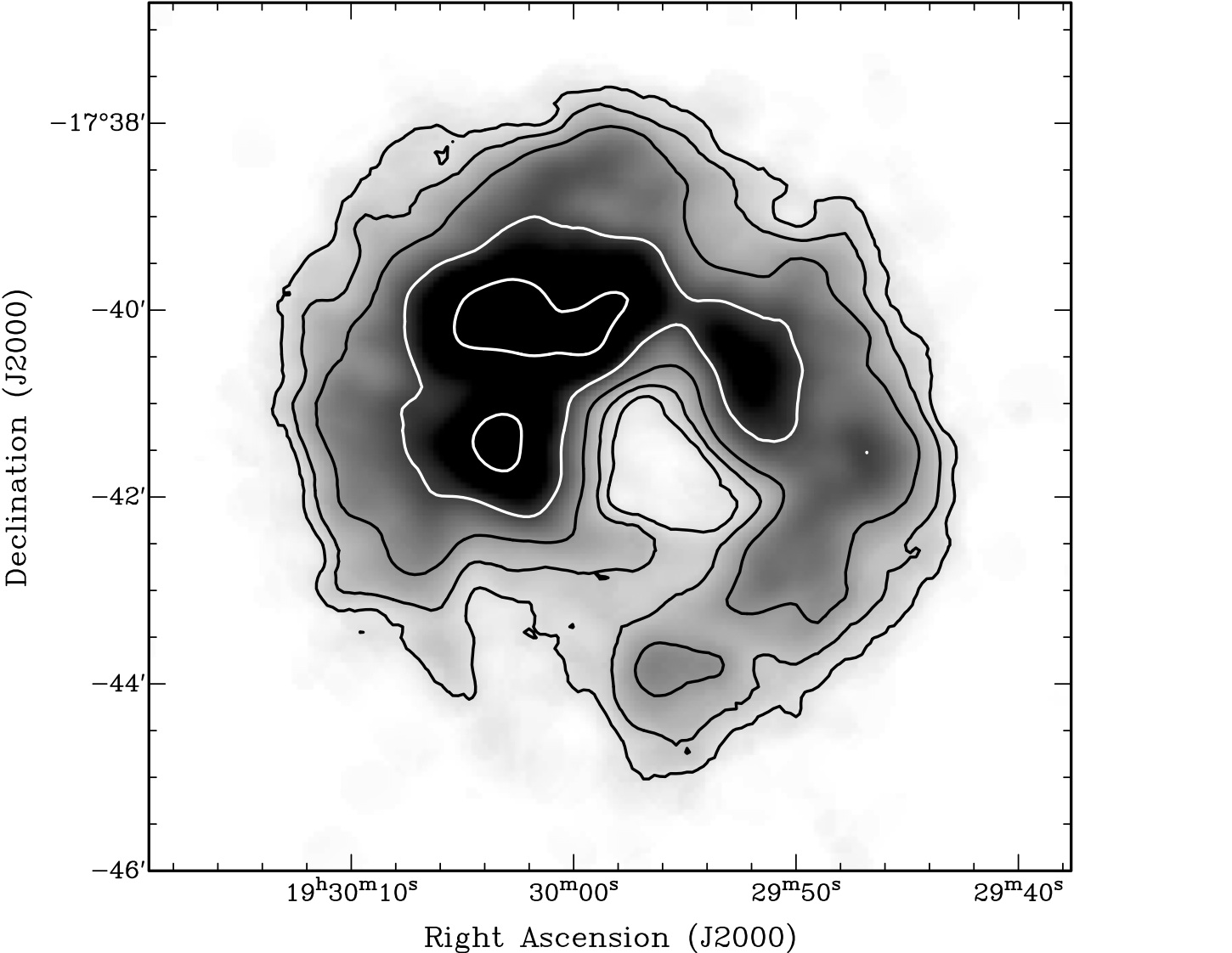}
   \includegraphics[width=\columnwidth]{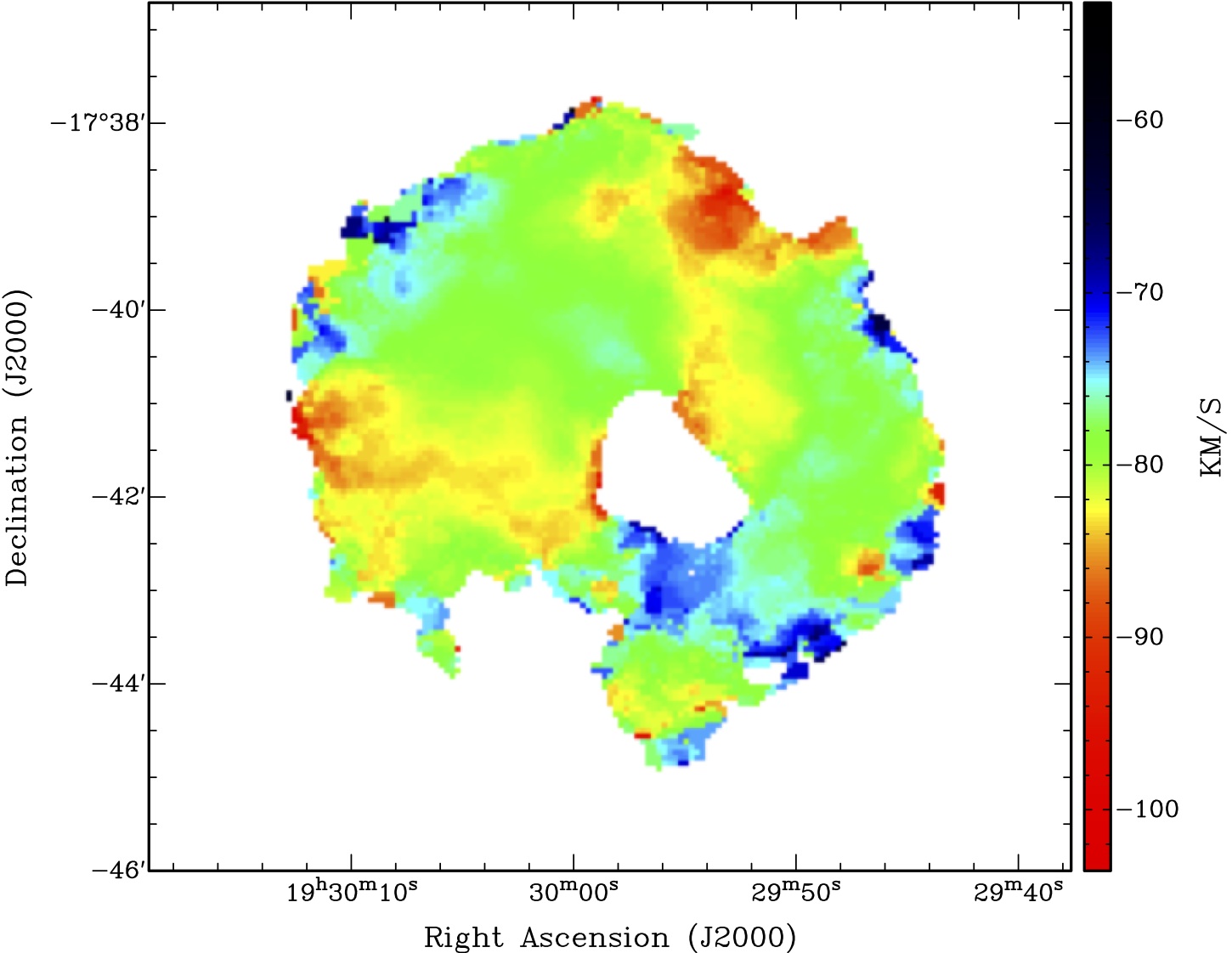}
     \caption{Total \HI\ map and velocity field obtained as a weighted average of the line profile. Both images have spatial resolution of $30\arcsec\times 30\arcsec$, the contours in the \HI\ maps are at 0.5, 1, 2, 4, 8 $\times10^{20}$~cm$^{-2}$.}
        \label{HImaps}
    \end{figure*}

We analysed the HI data cubes for Sgr~dIrr made publicly available from the LITTLE THINGS Survey \citep{littlethings}
using the Groningen Image Processing SYstem (GIPSY)\footnote{\tiny http://www.astro.rug.nl/~gipsy/}.
We first smoothed the HI data cube to a resolution of $30'' \times 30''$ and produced a new total \HI\ map and velocity field.
We built the total \HI\ map by adding up the emission after the application of a mask to each channel.
The mask was created by smoothing the original cube to a resolution of $60''$ and eliminating the regions with fluxes below $+2.5 \sigma_{60"}$, where $\sigma_{60"}$ is the r.m.s.\ noise of the smoothed cube.
We derived a total \HI\ mass of $\sim8\times 10^6~M_{\sun}$, in excellent agreement with what was found by \citet{littlethings} from the same data.
Fig.~\ref{HImaps} (left panel) shows the total HI map (contours) of Sgr~dIrr in greyscale and contours. The shape of the gas distribution is almost perfectly round with an inner depletion slightly off-set with respect to the optical centre (see also Fig.~\ref{maps6}).
This depletion may have been caused by intense star formation and be the sign of either gas removal or ionisation.

\begin{figure}
   \centering
   \includegraphics[width=0.95\columnwidth, angle=0]{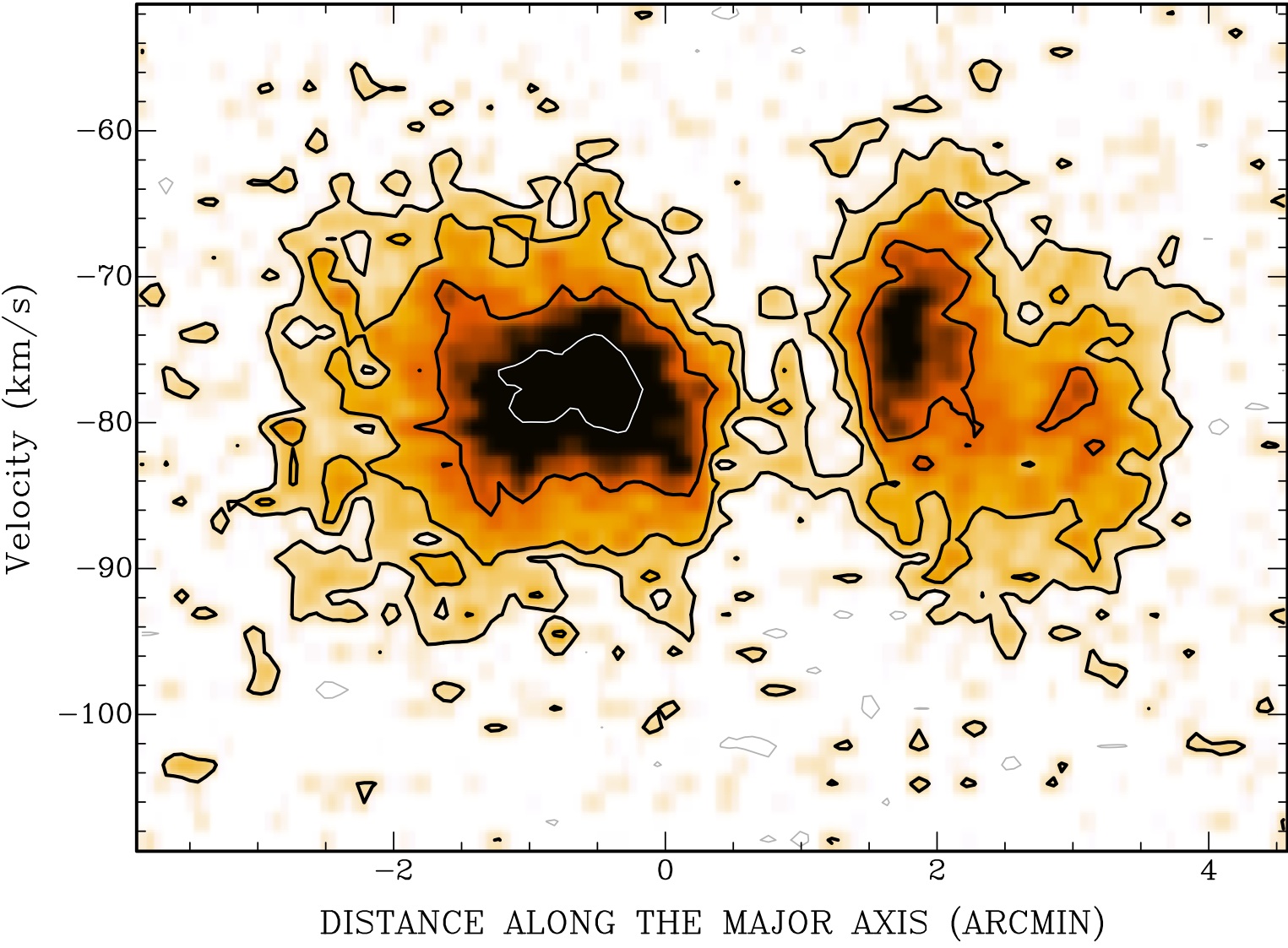}
     \caption{HI position-velocity diagram taken along the optical major axis of Sgr~dIrr, see Fig.\ \ref{maps6}, with position angle = 90$^{\circ}$. 
Contours are at -2, +2, 4, 8, 16 in units of r.m.s. noise, and the spatial resolution is about $22''$, as in Fig.\ \ref{maps6}.
The zero in x-axis corresponds to the optical centre of the galaxy, see Table \ref{Tab_par}}
        \label{HIpv}
    \end{figure}

The overall \HI\ kinematics can be appreciated from the velocity field in the right-hand panel of Fig.~\ref{HImaps}, derived using the first moment of the line-of-sight velocity distribution.
The kinematics is complex, and there seems to be no clear pattern of rotation or a systematic segregation of blue-shifted and red-shifted regions \citep[see also][]{younglo}.
To test the possible presence of low-level rotation, we extracted a position-velocity (p-v) diagram from the data cube along the major axis of the stellar distribution.
Figure~\ref{HIpv} shows this p-v plot and confirms the lack of any obvious velocity gradient.
Should one interpret the slight dislocation of the upper contours in Fig.~\ref{HIpv} as rotation, that would yield v$_{\rm rot} \sim 2 \times \frac{\sin(60)}{\sin(i)}$ km s$^{-1}$.
In an irregular system such as Sgr~dIrr, it is possible that the gas rotation is not aligned with the optical major axis \citep[e.g.,][]{Lelli+14}.
Thus, we checked for rotation along other directions but found only a slight gradient for angles close to the optical minor axis that is more likely associated with outflow/inflow motions of the gas (Battaglia \& Fraternali, in prep.).

The lack of a rotational signal in Sgr~dIrr can have different interpretations.
One possibility is that a rotation of a few km s$^{-1}$ may go undetected because it is lower than the \HI\ velocity dispersion, ranging from $\sim$5 to more than 10 km s$^{-1}$.
The \HI\ gas may also have a somewhat spherical and potentially pressure-supported distribution or, if a disk is present, it may have an inclination close to face-on, compatible with the projected round shape, which would make any line-of-sight rotational component very weak.
Finally, recent episodes of star formation may locally increase the non-circular motions of the \HI\ gas and make the signature of rotation temporarily undetectable.

The global \HI\ velocity profile of Sgr~dIrr can be fitted with a single Gaussian function with velocity dispersion $\sigma_{HI}= 8.6 \pm 0.1$~km\,s$^{-1}$.
In the inner regions, where the density is higher, the line profiles appear more complex, and a single Gaussian does not return a very good fit.
Considering two Gaussian functions, we find values of about 5 and 10 km\,s$^{-1}$ for the narrow and broad components, respectively.
This maybe be the same indication of two-phase interstellar medium as seen in other dwarf galaxies \citep[e.g.,][]{younglo, Ryan-Weber+08}, including VV124 (Paper~I).

Given the lack of rotation, it is difficult to reliably estimate the dynamical mass of Sgr~dIrr.
However, we can derive an order-of-magnitude estimation using the average observed \HI\ velocity dispersion $\sigma_{HI}\simeq 8.6$~km\,s$^{-1}$.
Assuming that the \HI\ has a spherical and isotropic distribution and it that is in hydrostatic equilibrium in the potential well of Sgr~dIrr, the dynamical mass \citep[$M_{\rm dyn}\sim 3 r_{\rm G} \sigma^2/ G$;][]{Ryan-Weber+08}\footnote{In Paper~I the rough estimate of the dynamical mass of VV124 was erroneously obtained with the same formula with a normalisation factor of 5 instead of 3. The guess estimates of the dynamical  mass and mass-to-light ratio reported in that paper must be therefore reduced by a factor of $\frac{3}{5}$.} would be about $5 \times 10^7 M_{\odot}$ for a gravitational radius $r_{\rm G}\sim r_{\rm HI} =$ 1 kpc.
This would lead to a mass-to-light ratio $M_{\rm dyn}/L_V\sim 9$, around the typical value for dwarfs of this luminosity (see M12, his Fig.~11, in particular).

\section{Summary and Discussion}
\label{disc}

We present new deep wide-field photometry of the isolated dwarf irregular galaxy Sagittarius. We do not see  any sign of a metallicity gradient in the old- and intermediate-age population of the galaxy, and we confirm that this population has a much more extended distribution than young stars (age$\la 1$~Gyr). 

Our data allowed us to trace the structure of the galaxy out to much greater distances from the centre ($r_{\epsilon}\simeq 5.0\arcmin$) and to much fainter surface brightness levels ($\mu_V\simeq 30.0$~mag/arcsec$^2$) than in any previous study.
The major axis profile is fully consistent with a purely exponential model from $r_{\epsilon}\simeq 200$~pc out to
the least measurable point, at $r_{\epsilon}\simeq 1600$~pc, and available surface photometry indicates a flattening in the innermost $\sim 200$~pc. 

Surface density maps show that the stellar body of the galaxy is significantly elliptical with a boxy inner region surrounded by an elongated halo. This is significantly different from the morphology of the neutral hydrogen distribution, which is remarkably circular, it is off-centred with respect to the stellar body and displays multiple density peaks and holes \citep[see also][]{younglo}.

The analysis of recent publicly available \HI\ data showed that, while large scale gradients are seen in the velocity field, no clear sign of systemic rotation can be identified. Both a rotation amplitude lower than the \HI\ velocity dispersion and chaotic motions induced by astrophysical mechanisms (e.g., star formation, gas in-outflows) probably contribute to masking the underlying rotation signal, if any. Also the merging with another dwarf may have played a role in producing the disturbed \HI\ morphology and velocity field \citep[see, e.g.,][and references therein]{deason,fatta}.

As a result, no robust estimate of the dynamical mass can be obtained from \HI\ data. 
In this context, our new photometry can provide the targets for a kinematic follow up on RGB stars covering the whole extended body of the galaxy \citep[as done, e.g., in][for VV124]{kv124}, aimed at obtaining an estimate of the mass relying on tracers whose motion only depends on the gravitational potential.

\begin{acknowledgements}
We are grateful to the referee, R.A. Ibata, for his useful suggestions.

M.B, F.F, and A.S. acknowledge the financial support from PRIN MIUR 2010-2011 project ``The
Chemical and Dynamical Evolution of the Milky Way and Local Group Galaxies'',
prot. 2010LY5N2T. 

G. Battaglia gratefully acknowledges support through a Marie-Curie action Intra European Fellowship, funded from the European Union Seventh Framework Programme (FP7/2007-2013) under Grant agreement number PIEF-GA-2010-274151, as well as the financial support by the Spanish Ministry of Economy 
and Competitiveness (MINECO) under the Ram\'on y Cajal Program 
(RYC-2012-11537).

This research made use of the SIMBAD database, operated at the CDS, Strasbourg, France.
This research made use of the NASA/IPAC Extragalactic Database (NED) which is operated by the Jet Propulsion Laboratory, California Institute of Technology, under contract with the National Aeronautics and Space Administration.

\end{acknowledgements}

\bibliographystyle{apj}



\end{document}